\documentclass[aps, nofootinbib, prd, superscriptaddress, tightenlines,
notitlepage, twocolumn, floatfix]{revtex4-1}
\usepackage{tabularx}
\usepackage{graphicx}
\usepackage{epsfig}
\usepackage{amssymb,bm,tensor}
\usepackage{textcomp}
\usepackage{xcolor}
\usepackage{amsmath}
\usepackage[varg]{txfonts}
\usepackage{enumerate}
\usepackage{mathtools}
\usepackage[normalem]{ulem}
\usepackage{bm} 
\usepackage[OT1]{fontenc}
\usepackage[colorlinks]{hyperref}
\usepackage{multirow}
\usepackage{makecell}
\usepackage{soul}
\newcommand{\rmd}{{\rm d}}
\newcommand{\Reply}[2]{{\color{red}\sout{#1}\color{black}{#2}}}

\newcommand{\KIAA}{\affiliation{Kavli Institute for Astronomy and
Astrophysics, Peking University, Beijing 100871, China}}
\newcommand{\DoA}{\affiliation{Department of Astronomy, School of Physics,
Peking University, Beijing 100871, China}}
\newcommand{\SOP}{\affiliation{School of Physics,
Peking University, Beijing 100871, China}}
\newcommand{\MPIfR}{\affiliation{Max-Planck-Institut f\"ur Radioastronomie,
Auf dem H\"ugel 69, D-53121 Bonn, Germany}}
\newcommand{\NAOC}{\affiliation{National Astronomical Observatories,
Chinese Academy of Sciences, Beijing 100012, China}}

\begin{document}

\title{Scalarized neutron stars in massive scalar-tensor gravity: X-ray pulsars
and tidal deformability}
\date{\today}
\author{Zexin Hu}\SOP
\author{Yong Gao}\email[Corresponding author: ]{gaoyong.physics@pku.edu.cn}\DoA\KIAA
\author{Rui Xu}\KIAA
\author{Lijing Shao}\email[Corresponding author: ]{lshao@pku.edu.cn}\KIAA\MPIfR\NAOC

\begin{abstract}
Neutron stars (NSs) in scalar-tensor theories of gravitation with the phenomenon
of spontaneous scalarization can develop significant deviations from general
relativity. Cases with a massless scalar were studied widely. Here we compare
the NS scalarizations in the Damour--Esposito-Far{\`e}se theory, the
Mendes-Ortiz  theory, and the $\xi$-theory with a massive scalar field.
Numerical solutions for slowly rotating NSs are obtained. They are used to
construct the X-ray pulse profiles of a pair of extended hot spots on the
surface of NSs.  We also calculate the tidal deformability for NSs with
spontaneous scalarization which is done for the first time with a massive scalar
field. We show the universal relation between the moment of inertia and the
tidal deformability.  The X-ray pulse profiles, the tidal deformability, and the
universal relation may help to constrain the massive scalar-tensor theories in
X-ray and gravitational-wave observations of NSs, including the Neutron star
Interior Composition Explorer (NICER) satellite, Square Kilometre Array (SKA)
telescope, and LIGO/Virgo/KAGRA laser interferometers.
\end{abstract}

\maketitle

\section{Introduction}
\label{sec:intro}

Scalar-tensor (ST) theories of gravitation are the simplest extension to general
relativity (GR) by adding real scalar fields which mediate gravity in
part~\cite{Damour:1992we}. The properties of these theories depend on how the
scalar fields couple with the metric and matters, as well as the potential term
of scalars in the action. It has been demonstrated that for some choices of the
coupling, the theories provide solutions which are close enough to GR in the
weak-field regime while predict large differences in the strong-field
regime~\cite{Damour:1993hw, Damour:1996ke, Mendes:2016fby, Arapoglu:2019mun}.
Therefore, it is particularly interesting to study those kinds of ST theories
since they provide explicit examples of large deviations from GR in the
non-perturbative regime while still satisfy all experimental and observational
bounds from weak-field experiments~\cite{Will:2018bme}. One well-studied set of
the ST theories with such properties are characterized by a phenomenon called
{\it spontaneous scalarization}~\cite{Damour:1993hw}. In this work we will focus
on neutron stars (NSs), which are strong-field systems available to
astrophysical observations and possess matters with extreme densities and
pressures~\cite{Freire:2012mg, Shao:2016ezh, Shao:2019gjj, Liu:2020hkx} which
are coupled to the scalar fields in the ST theories.

First discovered by Damour and Esposito-Far{\`e}se (DEF) in a mono-scalar ST
theory, spontaneous scalarization can lead to non-perturbative, order-of-unity
deviations from GR~\cite{Damour:1993hw}, as the scalar-field vacuum becomes
unstable to the growth of the field in the presence of highly compact objects
like NSs. We will focus on the specific class of ST theories where only one
scalar field exists, and we focus on investigations with NSs. As we discuss in
detail in Sec.~\ref{sec:theory}, spontaneous scalarization for NSs can also be
understood as a Landau's phase transition with the star's baryonic mass taken as
the control parameter~\cite{Damour:1996ke, EspositoFarese:2004cc,
Sennett:2017lcx}.  For a sufficiently compact NS, the solution with a nontrivial
scalar field is energetically favored than the unstable solution that is
composed of the GR metric and a vanishing scalar field. 

The main motivation to study scalarized NSs is that they provide a new avenue to
test gravity theories in the strong-field regime.  For asymmetric binary pulsar
systems, scalarized NSs can trigger the gravitational dipole radiation in ST
theories~\cite{Damour:1996ke, Freire:2012mg}.  The dipole radiation will make
the orbital energy loss faster than that in GR if there exists such an extra
scalar degree of freedom, which in turn can be constrained by pulsar timing
observations~\citep{Damour:1996ke}. Besides, gravitational waves (GWs) from
binary NS mergers have been directly detected with laser interferometers
\citep{LIGOScientific:2017vwq} and can be analyzed to test the dipole radiation
\citep{Will:1994fb, Barausse:2012da, LIGOScientific:2018dkp, Zhang:2017sym,
Zhao:2021bjw} and put constraints on ST theories.

Most early studies of the ST theories with spontaneous scalarization considered
a massless scalar field.  However, observations of the pulsar--white dwarf
orbital decay due to gravitational dipole radiation have constrained the 
parameter space of the massless ST theories stringently~\cite{Freire:2012mg,
Shao:2017gwu,  Anderson:2019eay, Zhao:2019suc, Guo:2021leu}. Moreover, with a
massless scalar field, the whole Universe could have been scalarized, which is
clearly ruled out by the combination of Solar-system and cosmological
observations~\cite{Anderson:2016aoi}.  A simple and natural way to avoid these
difficulties is to consider a massive scalar field rather than a massless
one~\cite{Damour:1996ke, Ramazanoglu:2016kul, Yazadjiev:2016pcb}. Due to the
mass term, the scalar field outside the star suffers a Yukawa-type suppression
such that the scalar field's contribution to the gravitational radiation can be
effectively effaced.  Therefore, much of the parameter space for massive ST
theories is still not excluded by the pulsar timing
observations~\cite{Alsing:2011er, Liu:2020moh}. Besides, even a very light mass
of the scalar field can prevent the scalarization of the Universe.

Different coupling functions between the scalar field and matters correspond to
different kinds of ST theories. Currently, the DEF
theory~\citep{Damour:1993hw,Damour:1996ke}, the Mendes-Ortiz (MO)
theory~\citep{Mendes:2014ufa,Mendes:2016fby}, and the
$\xi$-theory~\citep{Damour:1996ke, Xu:2020vbs} are commonly used and all of them
can lead to spontaneous scalarization of NSs.  The DEF theory provides the
simplest coupling function possessing the spontaneous scalarization effect, and 
it is the most commonly studied ST theory since the seminal work by Damour and
Esposito-Far{\`e}se~\cite{Damour:1993hw, Damour:1996ke}. The $\xi$-theory is a
well-motivated theory, which arises from inflationary models where $\xi$
represents a dimensionless coupling constant in the
Lagrangian~\cite{Salopek:1988qh}. However, the explicit form of the $\xi$-theory
in the Einstein frame is not convenient for calculation. Therefore Mendes and
Ortiz~\cite{Mendes:2016fby} constructed an analytical approximation of the
$\xi$-theory, which is the so-called MO theory.  After adding the mass term for
the scalar field, the phenomenon of spontaneous scalarization still exists for
all of those theories~\cite{Mendes:2016fby,Ramazanoglu:2016kul,Xu:2020vbs}. In
this work, we calculate and compare the structure of a slowly rotating NS, and
its macroscopic properties, including the mass-radius relation, the moment of
inertia, and the tidal deformability, in the DEF, MO, and $\xi$ theories. We
also give explanations to the common features shown in spontaneous
scalarization. The study serves as a comprehensive comparison between different
ST theories with a massive scalar field, and discusses the relevance to a couple
of current and near-future observations.

As we mentioned before, the massive ST theories are poorly constrained by the
pulsar--white dwarf binary systems \citep{Shao:2017gwu,  Anderson:2019eay,
Zhao:2019suc, Guo:2021leu} due to the Yukawa-type suppression of the scalar
field. However, the structure of scalarized NSs (e.g., mass and radius) in ST
theories can still be very different from that in GR. Therefore, one possible
method to constrain the parameter space of the massive ST theories is using the
mass-radius relation (see e.g., discussions in Sec.~7 of
Ref.~\cite{Hu:2020ubl}).  \Reply{}{Moreover, as discussed 
by Silva and Yunes~\cite{Silva:2019leq}, it is possible to perform strong-gravity tests with 
X-ray observations using a recently developed pulse profile model beyond 
general relativity. The
X-ray pulse profiles from hot spots on the surface of a NS will be different in
ST theories from those in GR because of both the difference in the mass-radius
relation and the difference in the lightlike geodesic~\cite{Silva:2018yxz, Sotani:2017rrt}.}

We expect the theories in
consideration to be constrained by the X-ray timing data from the ongoing
Neutron star Interior Composition Explorer (NICER) program (see e.g., recent
results in Refs.~\cite{Miller:2021qha, Raaijmakers:2021uju}), and future planned
X-ray satellites like THESEUS~\cite{Ciolfi:2021gzg}.  As an application of the
numerical results of the scalarized NSs, we follow
Refs.~\cite{Xu:2020vbs,Silva:2018yxz} to calculate the X-ray pulse profiles of
NSs with spontaneous scalarization and generalize their results to the case of
extended hot spots for massive ST theories. 

GW observations from binary NSs can also give constraints on massive ST
theories. For coalescing binary NSs, the size of the stars cannot be ignored at
the end of inspiral as they will be deformed in the tidal fields of their 
companions, and the deformation of each star contributes a correction to the
waveform's phasing~\cite{Hinderer:2007mb}. A remarkable GW discovery of a binary
NS system is GW170817 by the LIGO/Virgo detector network, where analysis has
been performed to extract a constraint for the tidal deformability from the
data~\cite{LIGOScientific:2017vwq, LIGOScientific:2018hze}. In
Ref.~\cite{Pani:2014jra}, the tidal deformability of NSs in massless DEF
theories is calculated, and it is shown that the deviation from GR can be large
in some parameter space. In this work, we calculate the tidal deformability of
NSs in massive ST theories which, to our knowledge, is done for the first time.

Discovered by Yagi and Yunes ~\cite{Yagi:2013awa,Yagi:2013bca}, the
dimensionless moment of inertia, the tidal deformability, and the spin-induced
quadrupole moment of slowly rotating NSs in GR satisfy nearly universal
relations which do not depend on the NSs' equation of state (EOS). It has been
found that in many other alternatives to GR the relations still hold with a high
accuracy.  The universal relations in massless ST theories have been shown by
\citet{Pani:2014jra}, and for the viable parameter space of the theories, the
deviation from GR is lower than 2\%. Nevertheless, the universal relations may
still be an interesting method to help us discriminate between GR and the ST
theories without the detailed knowledge of the realistic EOS of NSs. In this
work, we will discuss the universal relation between the moment of inertia and
the tidal deformability in massive ST theories for the first time.

The organization of the paper is as follows. In Sec.~\ref{sec:theory} we briefly
introduce the massive ST theories considered in this paper and give a discussion
about the spontaneous scalarization. The differential equations for a slowly
rotating NS are derived in Sec.~\ref{sec:NS} and the numerical solutions of NSs'
structure are also displayed there. As an application of the results, in
Sec.~\ref{sec:X-ray} we show the X-ray pulse profiles from a pair of hot spots
on the surface of a scalarized NS.  In Sec.~\ref{sec:tidal}, the tidal
deformability of a scalarized NS is calculated by considering a perturbation
both in the metric and the scalar field. Then, we show the relation between the
moment of inertia and the tidal deformability in Sec.~\ref{sec:I-lambda}.
Finally, we summarize the paper in Sec.~\ref{sec:summary}.

Throughout this work, we use the geometrized unit system where the ``bare''
gravitational constant $G$ and the speed of light $c$ are set to 1, except when
traditional units are written out explicitly. The convention of the metric is
$\left(-,+,+,+\right)$.

\section{Massive scalar-tensor theories}
\label{sec:theory}

The action of a massive ST theory in the Einstein frame
reads~\cite{Damour:1996ke, Damour:2007uf, Mendes:2014ufa},
\begin{align}
  \label{eqn:S}
  S=& \frac{1}{16\pi} \int \rmd^{4} x \sqrt{-g} \left[R-2 g^{\mu \nu}
  \partial_{\mu} \varphi \partial_{\nu}
  \varphi - V\left(\varphi\right) \right] \nonumber \\ 
  & +S_{\mathrm{m}}\left[\Psi_{\rm m}, A^{2}\left(\varphi\right) g_{\mu
  \nu} \right]  \,,
\end{align}
where $R$ is the curvature scalar constructed from the metric $g_{\mu\nu}$,
$\varphi$ is the scalar field coupled with the metric and matters, and $S_{\rm
m}$ is the action of conventional matters where $\Psi_{\rm m}$ represents matter
fields collectively. It is convenient to formulate the field equations via the
metric $g_{\mu\nu}$ in the Einstein frame, while matters are directly coupled to
the Jordan frame metric $\tilde g_{\mu\nu}$, which is related to the Einstein
metric $g_{\mu\nu}$ via a conformal transformation
$\tilde{g}_{\mu\nu}=A^2\left(\varphi\right)g_{\mu\nu}$. Jordan frame metric is
the one measured by experiments. For this reason, the Jordan frame is usually
referred to as the physical frame.

We take the scalar potential $V\left(\varphi\right)$ as 
\begin{equation}
  V\left(\varphi\right) = 2 m_\varphi^2 \varphi^2 \,,
  \label{scalarpotential}
\end{equation}
which represents a scalar field with a constant scalar mass $m_\varphi$ in the
Einstein frame. The corresponding Compton wavelength of the scalar field is
defined as 
\begin{equation}
  \lambda_{\varphi}=\frac{2\pi \hbar}{m_\varphi} \,,
\end{equation}
where $\hbar$ is the reduced Planck constant. We will use the reduced wavelength
${\lambdabar_{\varphi}} \equiv \lambda_{\varphi} /2\pi$ in the rest of
the work. 

By varying the total action in Eq.~(\ref{eqn:S}) with respect to the tensor
field $g_{\mu\nu}$ and the scalar field $\varphi$, one obtains the field
equations~\cite{Damour:1992we}
\begin{align}
  \label{eqn:R}
    R_{\mu\nu}=& 2\partial_{\mu}\varphi\partial_{\nu}\varphi+\frac{1}{2}g_{\mu\nu}V\left(\varphi\right)+
    8\pi\left(T_{\mu\nu}-\frac{1}{2}g_{\mu\nu}T\right) \,, \\
  \label{eqn:phi}
    \square\varphi=& \frac{1}{4}\frac{\rmd V\left(\varphi\right)}{\rmd \varphi}-4\pi\frac{\rmd \ln A\left(\varphi\right)}{\rmd \varphi}T \,.
\end{align}
Note that all tensorial operations in Eqs.~(\ref{eqn:R}--\ref{eqn:phi}) are
performed in the Einstein frame. $T_{\mu\nu}$ is the energy-momentum tensor for
conventional matters, which is defined as $T_{\mu\nu} \equiv -{2}{(-g)}^{-1/2}
\, {\delta S_{\rm m}}/{\delta g^{\mu\nu}}$.  While the energy-momentum tensor in
the physical frame can be represented as $\tilde{T}_{\mu\nu} \equiv
-{2}{({-\tilde{g}})}^{-1/2} \, {\delta S_{\rm m}}/{\delta \tilde{g}^{\mu\nu}}=
A^{-2}T_{\mu\nu}$, which satisfies the conservation equation
$\tilde{\nabla}_{\nu} \tilde{T}^{\mu \nu}=0$.  For an ideal fluid, the
energy-momentum tensor in the physical frame is
\begin{equation}
  \label{eqn:energy momentum}
  \tilde{T}^{\mu \nu}=\left(\tilde{\epsilon}+\tilde{p}\right) \tilde{u}^{\mu} \tilde{u}^{\nu}+\tilde{g}^{\mu \nu} 
  \tilde{p}\,.
\end{equation}

As  commonly used in the literature, we define  $\alpha\left(\varphi\right)
\equiv {\rmd \ln A\left(\varphi\right)}/{\rmd \varphi}$.  From
Eq.~(\ref{eqn:phi}), it is clear that $\alpha\left(\varphi\right)$ plays the
role of measuring the field-dependent coupling strength between the scalar field
and matters. Generally, we expand $\ln A\left(\varphi\right)$ around a
background scalar field $\varphi_0$,
\begin{equation}
  \ln A\left(\varphi\right)=\ln
  A\left(\varphi_0\right)+\alpha_0\left(\varphi-\varphi_0\right)
  +\frac{1}{2}\beta_0\left(\varphi-\varphi_0\right)^2 +\cdots \,,
\end{equation}
where $\varphi_0$ denotes the asymptotic value of $\varphi$ at spatial infinity.
In the theories with a massless scalar field, all the post-Newtonian parameters
are proportional to powers of $\alpha_0$~\cite{Damour:1993hw}, and the Solar
system experiments have constrained $\alpha_0$ very close to 
zero~\cite{Will:2018bme}, which indicates a very weak coupling between matters
and the scalar field in the weak-field regime. The quadratic coefficient
$\beta_0$ is nearly not constrained by Solar system experiments since it enters
in the post-Newtonian expansion with a multiplier of $\alpha_0^2$, and thus is
always suppressed by the already well constrained coefficient $\alpha_0$. In
this work, with a massive scalar, the asymptotic value $\varphi_0$ has to be
zero due to the scalar potential in Eq.~(\ref{scalarpotential}), and  we take
$A\left(\varphi_0\right)=1$ and $\alpha_0=0$ for simplicity.

Currently, there are three main representative forms for the function
$\alpha\left(\varphi\right)$ that can lead to spontaneous scalarization. We will
describe these theories as follows.

In the DEF theory~\cite{Damour:1993hw, Damour:1996ke}, the function form of the
conformal factor is
\begin{equation}\label{eq:Avarphi:DEF}
  A\left(\varphi\right)=e^{ \frac{1}{2} \beta \left(\varphi - \varphi_0\right)^{2} }\,,
\end{equation}
which provides the coupling function 
\begin{equation}
  \alpha\left(\varphi\right) = \beta \left(\varphi - \varphi_0\right) \,.
\end{equation}
This coupling function is linear in $\varphi$ and can give large deviations from
GR in the strong-field regime due to the positive correlation between the
coupling strength and the scalar field. 

A well-motivated form of $A\left(\varphi\right)$ from cosmology is the standard
non-minimal coupling~\cite{Damour:1996ke, Arapoglu:2019mun,Xu:2020vbs}
\begin{equation}\label{eq:Avarphi:xi}
  A\left(\varphi\right) = \frac{1}{\sqrt{1+\xi \Phi\left(\varphi\right)^2}} \,,
\end{equation}
where $\Phi\left(\varphi\right)$ is the scalar field in the physical frame and implicitly given by
\begin{align}
  \label{eqn:Phi2phi}
  \varphi =& \varphi_{0} +  \frac{\chi}{2 \sqrt{2} \xi} \ln
  \left[1+2 \chi \Phi\left(\sqrt{1+\chi^{2} \Phi^{2}}+\chi \Phi\right)
  \right] \nonumber \\
  & + \frac{\sqrt{3}}{2} \ln \left[1-2 \sqrt{6} \xi \Phi \frac{\sqrt{1+\chi^{2}
  \Phi^{2}}-\sqrt{6} \xi \Phi}{1+\xi \Phi^{2}}\right] \,,
\end{align}
with $\chi \equiv \sqrt{\xi\left(1+6 \xi\right)}$, and $\xi>0$. The coupling
function of the $\xi$-theory, 
\begin{equation}
  \alpha\left(\varphi\right) = - \frac{\sqrt{2} \xi \Phi}{\sqrt{1+\xi\left(1+6 \xi\right)
  \Phi^{2}}}\,,
\end{equation}
cannot be written in terms of $\varphi$ in a closed form, so the calculation has
to be done partly in the physical frame with the scalar field described by
$\Phi$.

The MO theory~\cite{Mendes:2016fby} is an analytical approximation of the
$\xi$-theory. The function $A\left(\varphi\right)$ has the form
\begin{equation} \label{eq:Avarphi:MO}
  A\left(\varphi\right)=\left[\cosh \left(\sqrt{3} \beta \left(\varphi - \varphi_0\right)
  \right)\right]^{ \frac{1} {3 \beta}} \,,
\end{equation}
and the coupling function $\alpha\left(\varphi\right)$ is
\begin{equation}
  \alpha\left(\varphi\right)=\frac{1}{\sqrt{3}} \tanh \left[\sqrt{3} \beta
  \left(\varphi-\varphi_{0}\right)\right] \,.
\end{equation}

When $\beta=-2\xi$, the three classes of theories in
Eqs.~(\ref{eq:Avarphi:DEF}), (\ref{eq:Avarphi:xi}), and~(\ref{eq:Avarphi:MO})
hold some common properties.  First, the MO theory can be well approximated by
the $\xi$-theory. Moreover, the coupling functions in the three classes of
theories have the same expansion to the linear order of $\varphi$,
\begin{equation}
  \alpha\left(\varphi\right) \approx \beta\varphi+O\left(\varphi^2\right)
  \approx -2\xi\varphi+O\left(\varphi^2\right)\,,
\end{equation}
where we have used $\varphi_0=0$. This means that in the limit of $\varphi \to
0$, behaviors of these theories should be the same. 

\begin{figure}
  \centering 
  \includegraphics[width=7cm]{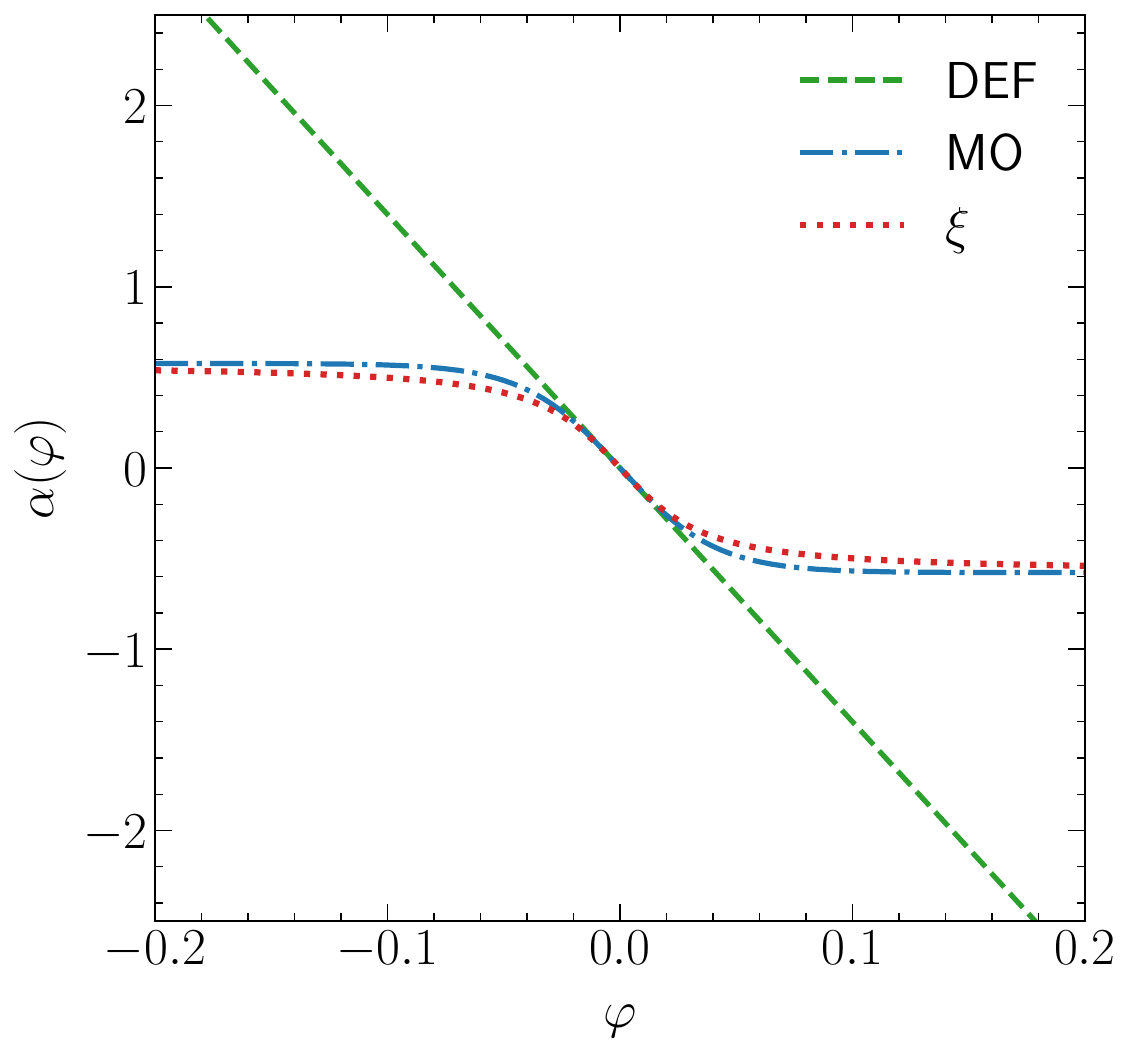}
  \caption{Coupling function $\alpha(\varphi)$ for three different theories. We
  have used $\beta=-2\xi=-14$.  }
  \label{fig:alpha-phi}
\end{figure}

To show the features of these theories, we plot the  coupling function
$\alpha\left(\varphi\right)$ in Fig.~\ref{fig:alpha-phi}. It is clear that the
DEF theory has the strongest coupling, and that the MO theory is a good
approximation of the $\xi$-theory though the coupling in the MO theory is always
a little stronger than that in the $\xi$-theory. Especially, the values of
$\alpha\left(\varphi\right)$ when $\varphi \rightarrow \infty$ are different in
the $\xi$ and the MO theories. But they become equal when $\xi$ goes to
infinity.

As show in Refs.~\cite{Damour:1993hw,Damour:1996ke,Mendes:2016fby}, for suitable
choices of $\beta$, all these three ST theories can have nonperturbative
strong-field deviations from GR, namely the spontaneous scalarization of NSs.
This name comes from the analogy to the phenomenon of spontaneous magnetization
of ferromagnet~\cite{Damour:1996ke}. The order parameter chosen by Damour and
Esposito-Far{\`e}se is the scalar charge $\mathcal Q$ developed by the NS, which
is defined via $\varphi\left(r\right)=\varphi_0+ {\mathcal
Q}/{r}+O\big(r^{-2}\big)$ at $r \to \infty$ in the massless DEF theory. This
parameter is undefined in massive ST theories due to the exponential Yukawa
factor in the asymptotic expansion of the scalar field. But one can take other
parameters, for example, the value of the scalar field at the center of the
star, as the order parameter and perform the same analysis. 

In the absence of the external field $\varphi_0$, one can take the star's energy
in the form of the usual Landau ansatz near the critical transition point, say,
$\mu\left(\mathcal Q\right)= a \left(\bar{m}_{\rm cr}-\bar{m}\right) {\mathcal
Q}^2 / 2 +  b {\mathcal Q}^4 / 4$, as a simple model exhibiting spontaneous
scalarization, where $\bar{m}$ is the baryon mass of the star.  The energy $\mu$
is minimal at the trivial solution $\mathcal Q=0$ when $\bar{m}<\bar{m}_{\rm
cr}$, while two energetically favored nontrivial solutions appear when
$\bar{m}>\bar{m}_{\rm cr}$ (cf. the Higgs mechanism in the Standard Model of
particle physics). Although $\mathcal Q=0$ is still a solution when
$\bar{m}>\bar{m}_{\rm cr}$, it is unstable to any perturbation of the scalar
field.

Another way to understand this phenomenon is to consider the instability of the
perturbation in the scalar field ~\cite{Ramazanoglu:2016kul}. To see that, we
expand Eqs.~(\ref{eqn:R}--\ref{eqn:phi}) to the linear order in $\varphi$,
\begin{align}
  R_{\mu\nu}=& 8\pi\left(\tilde{T}_{\mu\nu}-\frac{1}{2}g_{\mu\nu}\tilde{T}\right) \,, \\
  \label{eqn:linear phi}
  \square \varphi=& \left(m_\varphi^2-4\pi \tilde{T}\beta\right)\varphi \,.
\end{align}
Note that the equation for the metric is irrelevant to the perturbation in the
scalar field $\varphi$, so we can consider the stability of Eq.~(\ref{eqn:linear
phi}) in a background metric of the GR solution.

For matters described by a perfect fluid, we have
$\tilde{T}=-\tilde{\epsilon}+3\tilde{p}$, which is negative for ordinary
matters. So the right hand side in Eq.~(\ref{eqn:linear phi}) can be negative
for a sufficiently negative $\beta$ and the scalar field thus suffers an
instability (i.e., the tachyonic instability). It also shows that the mass term
of the scalar field suppresses the existence of spontaneous
scalarization~\cite{Ramazanoglu:2016kul}. \Reply{}{For highly compact NSs with some EOSs, the condition
that $\tilde{T}<0$ may not hold in the center of NSs~\cite{Podkowka:2018gib}.}  Therefore, a NS with a
large baryonic mass will not have spontaneous scalarization when $\beta<0$.  But
interestingly, the scalar field may suffer an instability when $\tilde{T}>0$ and
$\beta>0$ \citep{Mendes:2016fby}. In this work, we only concentrate ourselves on
the $\beta<0$ case.

\section{Neutron star structures in massive scalar-tensor theories}
\label{sec:NS}

In this section, we derive NS structures in massive ST theories.  Following
Refs.~\cite{Hartle:1967he,Hartle:1968si}, we consider a slowly rotating NS.
Keeping only the first order of the angular frequency $\Omega$, the stationary
and axisymmetric metric takes the form 
\begin{eqnarray}
  \label{eqn:rotation_metric}
  \rmd s^{2}&=&g_{\mu \nu} \rmd x^{\mu} \rmd x^{\nu} \nonumber\\
  &=&-e^{\nu \left(\rho\right)} \rmd t^{2}+\left(1-\frac{2 m\left(\rho\right)}{\rho}\right)^{-1} 
  \rmd \rho^{2} +\rho^{2} \left( \rmd \theta^{2} + \sin^{2} \theta \, \rmd \phi^2 \right) \nonumber\\
  && + 2  \rho^{2} \sin ^{2} \theta \left(\omega\left(\rho,\theta\right)-\Omega\right)\rmd t \rmd \phi\,,
\end{eqnarray}
where $\omega\left(\rho,\theta\right)$ is a function at the order of $\Omega$.
As show by~\citet{Hartle:1967he}, the function $\omega\left(\rho,\theta\right)$
can be expanded as 
\begin{equation}
  \omega\left(\rho,\theta\right)=\sum_{l=1}^{\infty} \omega_l\left(\rho\right) \left( -\frac{1}{\sin \theta}\frac{\rmd P_l}{\rmd \theta}\right)\,,
\end{equation}
and the radial functions $\omega_l(\rho)$ satisfy
\begin{equation}
  \frac{1}{\rho^4} \frac{\rmd}{\rmd\rho}\left[\rho^4 j(\rho) \frac{\rmd\omega_l}{\rmd\rho}\right]+\left[
  \frac{4}{\rho}-e^{(\lambda-\nu)/2}\frac{l(l+1)-2}{\rho^2}\right]\omega_l=0\,,
\end{equation}
where
\begin{eqnarray}
  j(\rho)&=&e^{-(\nu+\lambda)/2}\,,\\
  e^{\lambda} &=&\left(1-\frac{2 m(\rho)}{\rho}\right)^{-1}\,.
\end{eqnarray}

At large $\rho$, $\omega_l$ has the form
\begin{equation}
  \omega_l \to {\rm const.}\times \rho^{-l-2}+{\rm const.}\times \rho^{l-1}\,.
\end{equation}
For the space to be flat at large $\rho$, the only term remaining is $l=1$, and
thus the function $\omega\left(\rho,\theta\right)$ is independent of $\theta$.

\begin{figure*}
  \centering 
  \includegraphics[width=12cm]{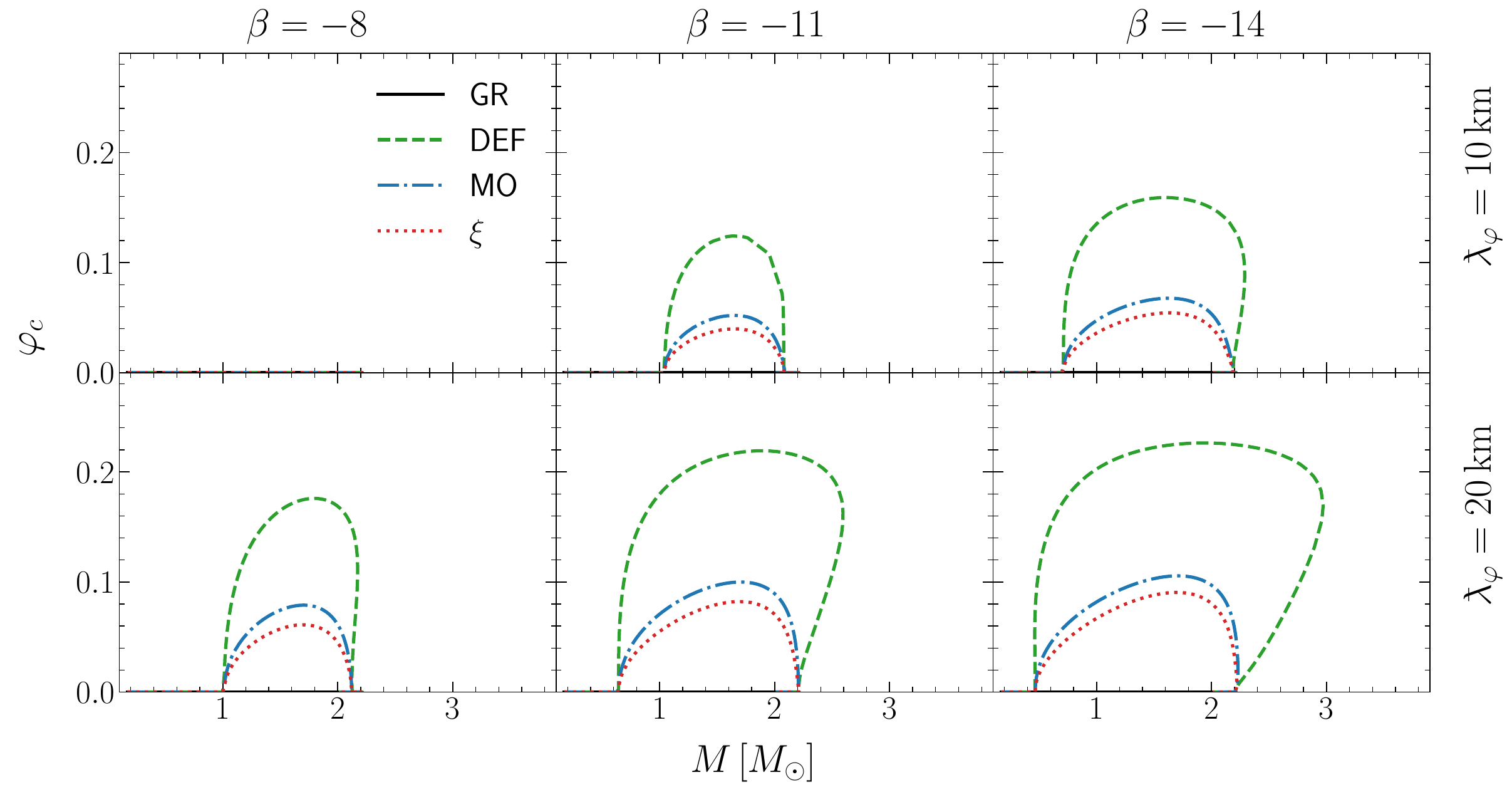}
  \caption{Spontaneous scalarization in the three theories, where $\varphi_c$ is
  the scalar field at the center of the NS and $M$ is the mass of the NS. EOS
  AP4 is used in the calculation. For large $M$, there are two solutions of
  $\varphi_c$, where the solution with the smaller $\varphi_c$ is
  unstable~\cite{Ramazanoglu:2016kul}. In the upper left panel, no spontaneous
  scalarization happens. \label{fig:phic-M}}
\end{figure*}

\begin{figure*}
  \centering 
  \includegraphics[width=12cm]{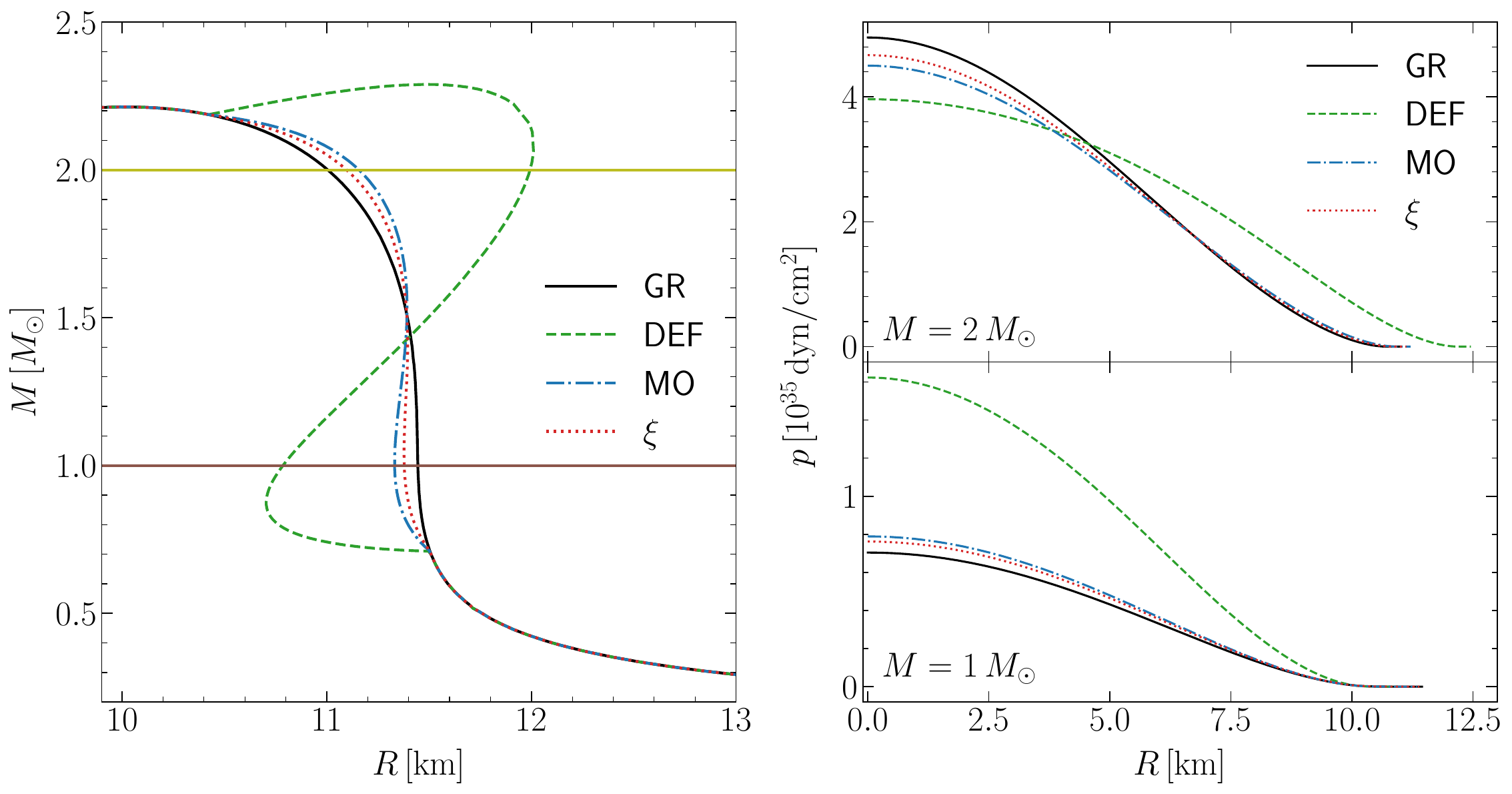}
  \caption{Mass-radius relation of NSs in GR and the ST theories with
  $\beta=-14$ and ${\lambdabar_{\varphi}}=10\,\rm km$ ({\it left}), and the pressure
  versus the radius of a NS with $M=1\,M_{\odot}$  ({\it lower right}) and
  $M=2\,M_{\odot}$  ({\it upper right}). EOS AP4 is used in the calculation.
  \Reply{}{The radius $R$ is defined by $R=A(\rho)\rho$, which is the length 
  in the physical frame.}
  \label{fig:M-R-P}}
\end{figure*}

Substituting Eqs.~(\ref{eqn:energy momentum}) and~(\ref{eqn:rotation_metric})
into Eqs.~(\ref{eqn:R}--\ref{eqn:phi}), one obtains systematic differential
equations for the metric and the scalar field~\cite{Damour:1992we,Damour:1996ke,
Xu:2020vbs,Ramazanoglu:2016kul},
\begin{eqnarray}
  \label{eqn:mprime}
  m^{\prime}&=&4 \pi \rho^{2} A^{4} \tilde{\epsilon}+\frac{1}{2} \rho\left(\rho-2 m\right) \varphi^{\prime 2}
  +\frac{1}{4} \rho^{2} V\,,\\
  \nu^{\prime}&=&\frac{8 \pi \rho^{2} A^{4} \tilde{p}}{\rho-2 m}+\frac{2 m}{\rho\left(\rho-2 m\right)}
  +\rho \varphi^{\prime 2}-\frac{1}{2} \frac{\rho^{2}}{\left(\rho-2 m\right)} V\,,\\
  \tilde{p}^{\prime}&=&-(\tilde{\epsilon}+\tilde{p})\left[\frac{4 \pi \rho^{2} A^{4} \tilde{p}}{\rho-2 m}
  +\frac{m}{\rho(\rho-2 m)}\right]\nonumber\\
  &&  - (\tilde{\epsilon}+\tilde{p})\left[\frac{1}{2} \rho \varphi^{\prime 2}+\frac{A^{\prime}}{A}-\frac{1}{4} \frac{\rho^{2}}{(\rho-2 m)}   
  V\right]\,,\label{eqn:pr}     \\
  \varphi^{\prime \prime}&=&\frac{4\pi \rho A^4}{\rho-2M}\left[\rho\left(\tilde{\epsilon}-\tilde{p}\right) 
  \varphi^{\prime}+\alpha \left(\tilde{\epsilon}-3\tilde{p}\right)\right]\nonumber\\
  &&-\frac{2\left(\rho -M\right)}{\rho\left(\rho-2M\right)}\varphi^{\prime}+\frac{\rho}{\rho-2M}\left(\frac{1}{2} \rho V 
  \varphi^{\prime}+\frac{1}{4}\frac{\rmd V}{\rmd \varphi}\right)\,,\\
  \label{eqn:omegaprime}
  \omega^{\prime \prime}&=&\frac{4 \pi \rho A^{4}}{\rho-2 m}\left(\tilde{\epsilon}+\tilde{p}\right)\left(\rho 
  \omega^{\prime}+4 \omega\right)+\left(\rho \varphi^{\prime 2}-\frac{4}{\rho}\right) \omega^{\prime}\,,
\end{eqnarray}
where the primes denote the derivatives with respect to $\rho$. 

\Reply{}{As mentioned before, for $\xi$-theory one needs to calculate with $\Phi$, the scalar field in the physical 
frame. Taking a derivative in Eq.~(\ref{eqn:Phi2phi}), one gets 
\begin{equation}
  \left(\frac{\rmd \varphi}{\rmd \Phi} \right)^2 = \frac{3}{4}\left(\frac{2\xi\Phi}{1+\xi\Phi^2}\right)^2+\frac{1}{2}\frac{1}{1+\xi\Phi^2}\,.
\end{equation}

Using this relation it is easy to change all the $\varphi'$, $\varphi''$, and $\rmd/\rmd \varphi$ appearing in 
Eqs.~(\ref{eqn:mprime}--\ref{eqn:omegaprime}) to $\Phi'$, $\Phi''$, and $\rmd/\rmd \Phi$; see Ref.~\cite{Xu:2020vbs} for more 
details.
}

To integrate the above equations, we start with the initial values at the center
of the NS,
\begin{eqnarray}
  m (\rho) &|_{\rho=0} = 0\,, \quad \nu (\rho) &|_{\rho=0} = 0\,, \quad \tilde{p} (\rho) |_{\rho=0} = \tilde{p}_c\nonumber\,,\\
  \varphi (\rho) &|_{\rho=0} = \varphi_c\,, \quad \varphi ^{\prime} (\rho) &|_{\rho=0} = 0
  \nonumber\,,   \\
  \omega (\rho)  &|_{\rho=0} = 1    \,, \quad \omega^{\prime} (\rho)  &|_{\rho=0} = 0 \,. 
\end{eqnarray}
Note that the initial value of $\nu$ is not equal to $0$ in the real case, one
needs to use the fact that the asymptotic value of $\nu$ at infinity is $0$ to
get the real initial value of $\nu$. The initial value of $\omega$ is also given
arbitrarily, since the equation for $\omega$ is homogeneous. Solving 
Eqs.~(\ref{eqn:mprime}--\ref{eqn:omegaprime}) with a given EOS, one gets the 
structure of a slowly rotating NS in the ST theory.  We take
AP4~\cite{Akmal:1998cf} as the example EOS to calculate NS structures in the
following.

\begin{table*}
  \renewcommand\arraystretch{1.3}
  \begin{center}
    \caption{NS structures in GR and the ST theories with EOS AP4.}
    \begin{tabular*}{\hsize}{@{}@{\extracolsep{\fill}}ccm{0.055\textwidth}<{\centering} m{0.055\textwidth}<{\centering}m{0.055\textwidth}<{\centering}m{0.055\textwidth}<{\centering} m{0.055\textwidth}<{\centering}m{0.055\textwidth}<{\centering}m{0.055\textwidth}<{\centering} m{0.055\textwidth}<{\centering}m{0.055\textwidth}<{\centering}m{0.055\textwidth}<{\centering} m{0.055\textwidth}<{\centering}m{0.055\textwidth}<{\centering}m{0.055\textwidth}<{\centering} @{}}
      \hline\hline
      \multicolumn{2}{c}{\multirow{2}{*}{Theory}}&\multirow{2}{*}{GR}&\multicolumn{3}{c}{$\beta=-11,\,\,{\lambdabar_{\varphi}}=10\,\rm km$}&\multicolumn{3}{c}{$\beta=-14,\,\,{\lambdabar_{\varphi}}=10\,\rm km$}&\multicolumn{3}{c}{$\beta=-11,\,\,{\lambdabar_{\varphi}}=20\,\rm km$}&\multicolumn{3}{c}{$\beta=-14,\,\,{\lambdabar_{\varphi}}=20\,\rm km$}\\
      &&&DEF& MO & $\xi$&DEF&MO&$\xi$&DEF&MO&$\xi$&DEF&MO&$\xi$\\
      \hline
      \multirow{3}{*}{$M=1\,M_{\odot}$}&$\varphi_C$&0&0&0&0&0.135&0.047&0.036&0.180&0.071&0.055&0.203&0.082&0.067\\
      &$R\,[\rm km]$&11.4&11.4&11.4&11.4&10.8&11.3&11.4&10.7&11.3&11.3&10.6&11.2&11.3\\
      &$I\,[\rm 10^{45}\,g\,cm^2]$&0.82&0.82&0.82&0.82&0.82&0.82&0.82&0.86&0.82&0.82&0.95&0.82&0.82\\ 
      \hline
      \multirow{3}{*}{$M=2\,M_{\odot}$}&$\varphi_C$&0&0.099&0.032&0.023&0.149&0.053&0.041&0.219&0.089&0.071&0.226&0.097&0.080\\
      &$R\,[\rm km]$&11.0&11.4&11.1&11.0&12.0&11.2&11.1&12.5&11.3&11.2&12.9&11.4&11.3\\
      &$I\,[\rm 10^{45}\,g\,cm^2]$&2.15&2.35&2.17&2.16&2.73&2.22&2.19&3.18&2.29&2.24&3.70&2.33&2.27\\
      \hline
    \end{tabular*}\label{tab:NSs}
  \end{center}
\end{table*}

For suitable choices of the parameters, namely the compactness of the NS and the
coupling strength, there exist solutions favouring a non-zero scalar field. In
Fig.~\ref{fig:phic-M} we show the values of the scalar field at the center of
the star versus the Arnowitt-Deser-Misner  (ADM) mass of the spacetime for
different choices of $\beta$ and ${\lambdabar_{\varphi}}$. The spontaneous
scalarization grows with both (the negative value of) $\beta$ and ${\lambdabar_
{\varphi}}$. A larger mass of the scalar field can suppress or even prevent the
happening of scalarization. Also we find that only those NSs with suitable
masses have nontrivial scalar fields.  NSs with too small or too large masses
only have the GR solution. These features are consistent with the analysis in
Sec.~\ref{sec:theory}.  Namely, as shown in Eq.~(\ref{eqn:linear phi}), the
scalar field suffers the instability only when $\big(m_\varphi^2-4\pi
\tilde{T}\beta\big)$ is sufficiently negative, which depends on both the scalar
mass and the coupling strength.  It is also found that, as expected, when
$\beta=-2\xi$, the DEF theory has the largest scalar field among these theories.
The MO theory has similar behaviours to the $\xi$-theory especially for large
values of $|\beta|$. 

Taking $\beta=-14$ and ${\lambdabar_{\varphi}}=10\,\rm km$ as an example, we
plot the mass-radius relation in the left panel of Fig.~\ref{fig:M-R-P}. The
results of different theories are qualitatively consistent with expectation.
With the appearance of the scalar field, the radius of the star with a small ADM
mass becomes smaller and vice versa. One can understand this feature by
comparing Eq.~(\ref{eqn:pr}) with the Tolman-Oppenheimer-Volkoff (TOV) equations
in GR,
\begin{align}
   \tilde{p}' =& -\left(\tilde{\epsilon}+\tilde{p}\right)\left(\frac{4 \pi \rho^{2} \tilde{p}}{\rho-2 m}
  +\frac{m}{\rho\left(\rho-2 m\right)}\right)\,, \\
  m'=& 4\pi \rho^2 \tilde{\epsilon}\,.
\end{align}
In Eq.~(\ref{eqn:pr}), the terms in the right-hand side with the factor
$A^4(\varphi)$ are suppressed by the appearance of the scalar field, which is
formally equal to replacing the constant $G$ by $A^4(\varphi) G$.  But the
scalar field itself provides extra gravitation via the terms in the second line
of Eq.~(\ref{eqn:pr}), as well as extra mass via the second and third terms in
the right-hand side of Eq.~(\ref{eqn:mprime}).

For an ordinary NS with spontaneous scalarization, the suppression of
$A^4(\varphi)$ and the extra gravitation provided by the scalar field have the
same order of effect inside the NS. As the mass of the NS increases, the
suppression effect of $A^4(\varphi)$ becomes more and more important due to the
increase of $\tilde{p}$ and $\tilde{\epsilon}$ inside the NS, while the extra
gravitation provided by the scalar field changes very slowly when the mass of
the NS is not close to the critical transition point of spontaneous
scalarization. Thus for a low-mass NS, the appearance of the scalar field makes
the radius of the star smaller, while a high-mass NS with spontaneous
scalarization has a larger radius. This is illustrated in the left panel of
Fig.~\ref{fig:M-R-P}. 

We also show the pressure $\tilde{p}$ as a function of $\rho$ for NSs whose
masses are $M=1\,{ M_\odot}$ and $M=2\, { M_\odot}$ respectively in the lower
and upper plots in the right panel of Fig.~\ref{fig:M-R-P}. For a low-mass NS
with spontaneous scalarization, its $\tilde{p}$ decreases faster than that in GR
due to the extra gravitation caused by the scalar field. For a high-mass NS, the
suppression effect of $A^4(\varphi)$ makes $\tilde{p}$ change slower.

\begin{figure}
  \centering 
  \includegraphics[width=6.5cm]{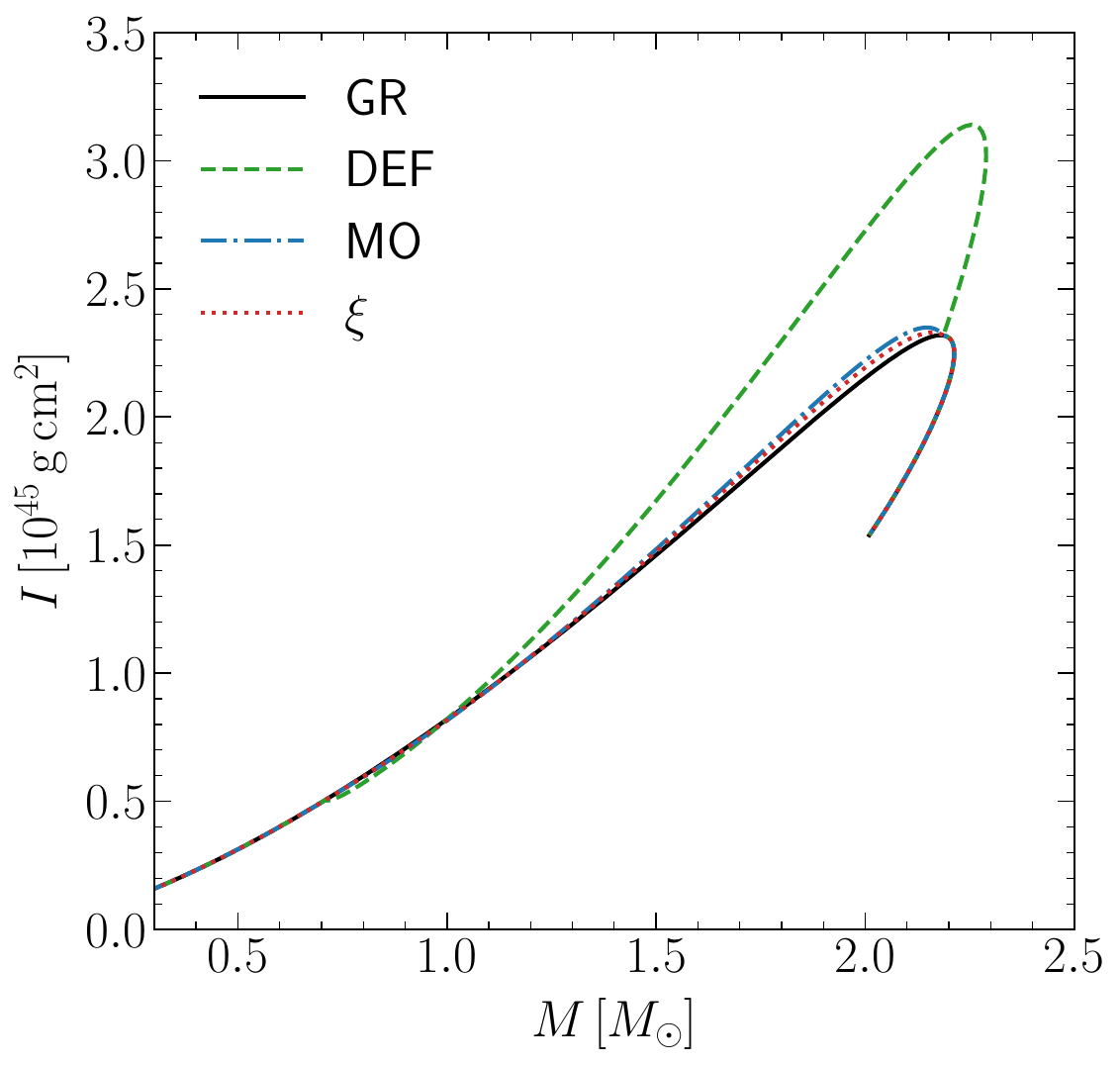}
  \caption{Moment of inertia $I$ versus the ADM mass $M$ with $\beta=-14$, ${
  \lambdabar_{\varphi}}=10\,\rm km$. EOS AP4 is used in the calculation.
  \label{fig:M-I}}
\end{figure}

In Fig.~\ref{fig:M-I} we show the relation between the ADM mass and the
moment of inertial $I$. The latter is defined by the asymptotic behaviour of the
metric element $g_{t \varphi}$. When $\rho \to \infty$, we have
\begin{equation}
  g_{t \varphi}=(\omega-\Omega)\rho^2 \sin^2 \theta \quad \to \quad -\frac{2J \sin^2 \theta}{\rho}\,.
\end{equation}
Thus the moment of inertia $I$ can be calculated via 
\begin{equation}
  I=\frac{J}{\Omega}= \left. \frac{1}{6}\frac{\rho^4 \omega'}{\omega} \right|_{\rho \to \infty}\,.
\end{equation}
Note that though in the mass-radius relation, a NS with spontaneous
scalarization could have a smaller or larger radius than that in GR, the moment
of inertia of the NS in the ST theories are almost always larger. 

Finally, in Table~\ref{tab:NSs} we list representative NS parameters in GR and
ST theories. One can notice the tend of the increasing scalarization as $\beta$
and $m_{\varphi}$ decrease. Generically speaking, the parameter space for
massive ST theories is less constrained than that in the massless
case~\cite{Shao:2017gwu, Zhao:2019suc, Anderson:2019eay, Guo:2021leu}, in
particular that a rather negative $\beta$ is still allowed by current bounds.
Therefore, NSs in massive ST theories can still develop quite different
properties from their GR solutions. From the observational aspects, significant
changes in the NS radius can be captured by X-ray timing satellite missions like
the NICER~\cite{Miller:2021qha,Raaijmakers:2021uju}, while significant changes
in the NS moment of inertia can be constrained by the Double Pulsar binary
system via, say, the Lense-Thirring effect, in the era of the Square Kilometre
Array (SKA)~\cite{Hu:2020ubl}. The DEF theory has particularly large effects to
be probed.

\section{X-ray pulsar pulse profiles}
\label{sec:X-ray}

We will discuss the X-ray pulsar pulse profiles in detail in this section.  As
an important goal of the hard X-ray timing instrument (e.g., NICER), using the
pulse profile observations from the hot spots on NS surfaces could help people
to measure the masses and radii of NSs to an accuracy of a few
percents~\cite{Watts:2016uzu,Miller:2021qha,Raaijmakers:2021uju}, which can be
used to determinate the EOS of the NSs. As show in Fig.~\ref{fig:M-R-P}, for a
given EOS, the mass-radius relation for scalarized NSs can have a large
deviation from that in GR. Consequently, The mass-radius relation of NSs can
also be used to constrain the parameter space of massive ST theories. In the
meanwhile, the appearance of the scalar field also affects the pulse profiles.
In this section, we follow Refs.~\cite{Silva:2018yxz,Xu:2020vbs} and calculate
the pulse profiles from a pair of extended hot spots on the surface of a NS with
spontaneous scalarization.

Following the assumptions in Refs.~\cite{Bogdanov:2006zd, Bogdanov:2008qm}, we
consider a model with a pair of hot spots sitting oppositely on the surface of a
slowly rotating NS. For simplicity, the metric outside the NS is taken to be
spherically symmetric. We neglect the effects of rotation and the deformations
of the NS on the spacetime, which have been shown to be small in GR for a slowly
rotating NS~\cite{Silva:2018yxz}.  We also assume that the distance between the
NS and the observer is large enough so that it can be treated mathematically as
infinity.

The geodesic equation for photons holds in the physical frame, so we need to use
the physical metric $\tilde{g}_{\mu\nu}=A^2(\varphi)g_{\mu\nu}$ to compute
photon trajectories. We define a new radial coordinate $r=A(\varphi)\rho$,
and the metric outside the star can be written as
\begin{equation}
  \rmd\tilde{s}^2=-g(r)\rmd t^2+f(r) \rmd r^2+r^2\rmd\theta^2+r^2\sin^2\theta \rmd\phi^2\,,
\end{equation}
with the coefficients
\begin{equation}
  f=\left(\frac{\rmd \rho}{\rmd r}\right)^2 A^2 \left(1-\frac{2 m(\rho)}{\rho}\right)^{-1},\quad g=A^2 e^{\nu}\,.
\end{equation}

As shown in Ref.~\cite{Xu:2020vbs}, for a point-like hot spot, the observed flux
can be expressed as 
\begin{equation}
  F(\nu)=g(R)\delta^4 I'(\nu',\alpha') \frac{\sin \alpha \cos\alpha}{\sin\psi} \frac{\rmd \alpha}{\rmd \psi} 
  \frac{\rmd S'}{D^2}\,,\label{eqn:flux}
\end{equation}
where $R$ is the radius of the star, $\delta$ is the Doppler factor caused by
the rotation, $I'(\nu', \alpha')$ is the radiation intensity at emission,
$\alpha$ represents the emission angle of the photons with respect to the local
radial direction, $\rmd S'$ is the proper differential area where the photons
are emitted, $D$ is the distance to the star and all primed quantities are
measured in the local rest frame on the surface of the NS (see Fig.~8 in
Ref.~\cite{Xu:2020vbs} for illustration).

If photons emitted at an angle $\alpha$ eventually propagate along the line of
sight and thus are observed, the angle $\psi$, spanning from the local radial
direction to the line of sight, must be related to the emission angle $\alpha$
via
\begin{equation}
  \psi=\sigma \int_{R}^{\infty} \frac{fg}{r^2} \frac{1}{\sqrt{1- {\sigma^2 g}/{r^2}}}\rmd r\,,
\end{equation}
where $\sigma \equiv {R} \sin\alpha / {\sqrt{g(R)}} $ is the impact parameter. 

The relative time delay can be defined by 
\begin{eqnarray}
  \delta t &=& t-t_0-\int_R^{\infty} \sqrt{\frac{f}{g}} \rmd r \nonumber\\
  &=& \int_R^{\infty} \sqrt{\frac{f}{g}} \left(\frac{1}{\sqrt{1- {\sigma^2 g}/{r^2}}}-1\right)\rmd r\,,\label{eqn:time delay}
\end{eqnarray}
where $t$ is the observation time and $t_{0}$ is the emission time. 

To consider a more realistic model, we extend the point-like hot spots in
Refs.~\cite{Silva:2018yxz, Xu:2020vbs} to finite regions. Following the method
in Ref.~\cite{Bogdanov:2008qm}, we put a grid of emission spots across the
stellar surface to construct X-ray emission regions and keep the radiation
intensity the same as the point-like spots.  Then, using geometric relations one
can calculate the observed angle $\psi$ in terms of the location of the
observer, the locations of the hot spots, and the star's rotating phase. The
observation time is related to $\psi$ by Eq.~(\ref{eqn:time delay}).  For each
infinitesimal hot spot one calculates the flux by Eq.~(\ref{eqn:flux}) and sums
them up to get the total flux received at a certain observation time.

\begin{figure*}
  \centering 
  \includegraphics[width=14cm]{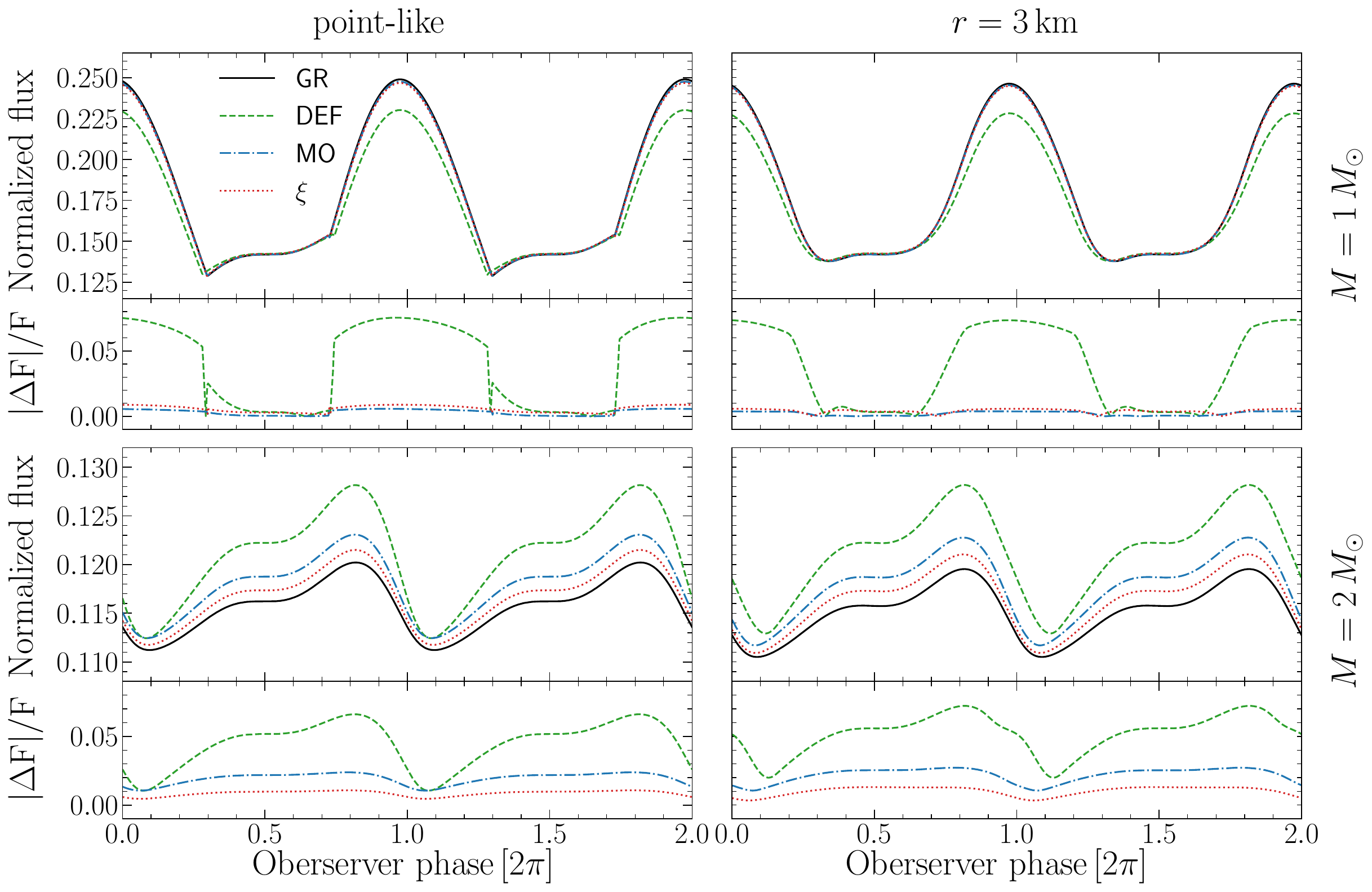}
  \caption{X-ray pulse profiles for NSs with isotropic emission intensity
  $I^{\prime}$. The \Reply{}{left column} is the results from point-like hot spots while the
  \Reply{}{right column} is the results from extended hot spots with spherical radius $r=3 \, \rm km$. The
  masses of the NSs are chosen as $1\,M_{\odot}$ for the upper set of curves and
  $2\,M_{\odot}$ for the lower set of curves. \Reply{}{The percental deviations from GR are plotted.} The integral radiating intensities
  of hot spots on the surface of NSs are chosen to be the same for the two cases. The angle
  between the NS spin axis and the hot spot on the northern hemisphere is
  $\pi/6$ and the spin axis is at an angle $\pi/3$ with respect to the line of
  sight. \Reply{}{The rotating frequency $f$ of the pulsar is assumed to be 
  $250 \, \rm Hz$} .We have used EOS AP4 in the calculation.  \label{fig:T-F} }
\end{figure*}

\begin{table*}
  \renewcommand\arraystretch{1.3}
  \begin{center}
    \caption{X-ray pulse profiles for NSs in GR and the ST theories. Results are
    only shown for the model with the isotropic radiation intensity and
    point-like hot spots. We have used EOS AP4 in the calculation.}
    \begin{tabular*}{\hsize}{@{}@{\extracolsep{\fill}}ccm{0.05\textwidth}<{\centering} m{0.05\textwidth}<{\centering}m{0.05\textwidth}<{\centering}m{0.05\textwidth}<{\centering} m{0.05\textwidth}<{\centering}m{0.05\textwidth}<{\centering}m{0.05\textwidth}<{\centering} m{0.05\textwidth}<{\centering}m{0.05\textwidth}<{\centering}m{0.05\textwidth}<{\centering} m{0.05\textwidth}<{\centering}m{0.05\textwidth}<{\centering}m{0.05\textwidth}<{\centering} @{}}
      \hline\hline
      \multicolumn{2}{c}{\multirow{2}{*}{Theory}}&\multirow{2}{*}{GR}&\multicolumn{3}{c}{$\beta=-11,\,\,{\lambdabar_{\varphi}}=10\,\rm km$}&\multicolumn{3}{c}{$\beta=-14,\,\,{\lambdabar_{\varphi}}=10\,\rm km$}&\multicolumn{3}{c}{$\beta=-11,\,\,{\lambdabar_{\varphi}}=20\,\rm km$}&\multicolumn{3}{c}{$\beta=-14,\,\,{\lambdabar_{\varphi}}=20\,\rm km$}\\
      &&&DEF& MO & $\xi$&DEF&MO&$\xi$&DEF&MO&$\xi$&DEF&MO&$\xi$\\
      \hline
      \multirow{4}{*}{$M=1\,M_{\odot}$}&peak flux&0.249&0.249&0.249&0.249&0.230&0.248&0.247&0.219&0.244&0.246&0.202&0.241&0.242\\
      &flux deviation (\%)&0&0.000&0.000&0.000&7.53&0.569&0.891&12.2&1.84&1.36&19.0&3.23&2.63\\
      &peak phase ($2\pi$)&0.976&0.976&0.976&0.976&0.976&0.976&0.976&0.976&0.976&0.976&0.974&0.976&0.976\\
      &phase deviation (\%)&0&0.00&0.00&0.00&0.00&0.00&0.00&0.00&0.00&0.00&0.18&0.00&0.00\\
      \hline
      \multirow{4}{*}{$M=2\,M_{\odot}$}&peak flux&0.120&0.125&0.122&0.120&0.128&0.123&0.121&0.126&0.124&0.122&0.121&0.125&0.124\\
      &flux deviation (\%)&0&4.67&1.70&0.309&7.03&2.78&1.47&5.14&3.29&2.28&1.30&4.19&3.30\\
      &peak phase ($2\pi$)&0.820&0.814&0.816&0.820&0.818&0.816&0.818&0.818&0.814&0.816&0.823&0.813&0.814\\
      &phase deviation (\%)&0&0.64&0.43&0.00&0.21&0.43&0.21&0.21&0.64&0.43&0.43&0.85&0.64\\
      \hline
    \end{tabular*}\label{tab:x-ray}
  \end{center}
\end{table*}

From the mass-radius relation shown in Fig.~\ref{fig:M-R-P}, we find that the
scalarized NSs have the largest deviation from GR in the radii around $M =
1.0\,M_{\odot}$ and $M = 2.0\,M_{\odot}$. Thus in Fig.~\ref{fig:T-F}, we plot
the X-ray pulsar profiles for two NSs with ADM masses $M = 1.0\,M_{\odot}$
(upper set of curves) and $M = 2.0\,M_{\odot}$ (lower set of curves). We take
the angle between the NS spin axis and the hot spots to be $\pi/6$ and the spin
axis is at an angle $\pi/3$ to the observer. The \Reply{}{left column} in this figure is
for the model with point-like hot spots and the \Reply{}{right column} is for the model
with extended circular emission regions of \Reply{}{$3\,\rm km$} radius. 

One can notice that for NSs with spontaneous scalarization, the phases of the 
profiles are almost the same as that in GR, which means that the time delay of
the photons in the ST theories is almost the same as that in GR. It is
reasonable because the scalar field suffers an exponential suppression with a 
characteristic length of ${\lambdabar_{\varphi}}$ outside the star and the
metric quickly becomes equal to that of GR. In the numerical results shown in
Fig.~\ref{fig:T-F}, the length ${\lambdabar_{\varphi}}$ is chosen to be $10\,\rm
km$, which is about the same as the NS's radius. The region with remarkable
values of the scalar field is only inside the NS.  The main difference of the
X-ray profiles in Fig.~\ref{fig:T-F} is the amplitudes of the fluxes, which have
noticeable deviations from GR around the peaks of the profiles. This feature is
mainly related to the difference in the mass-radius relation and dominated by 
the gravitational redshift. According to Fig.~\ref{fig:T-F}, the difference
between the point-like model and the extended model is relatively small.
Therefore, for a hot spot with radius about \Reply{}{$3\,\rm km$}, the point-like model is a
good approximation. 

\begin{figure}
  \centering 
  \includegraphics[width=0.9\linewidth]{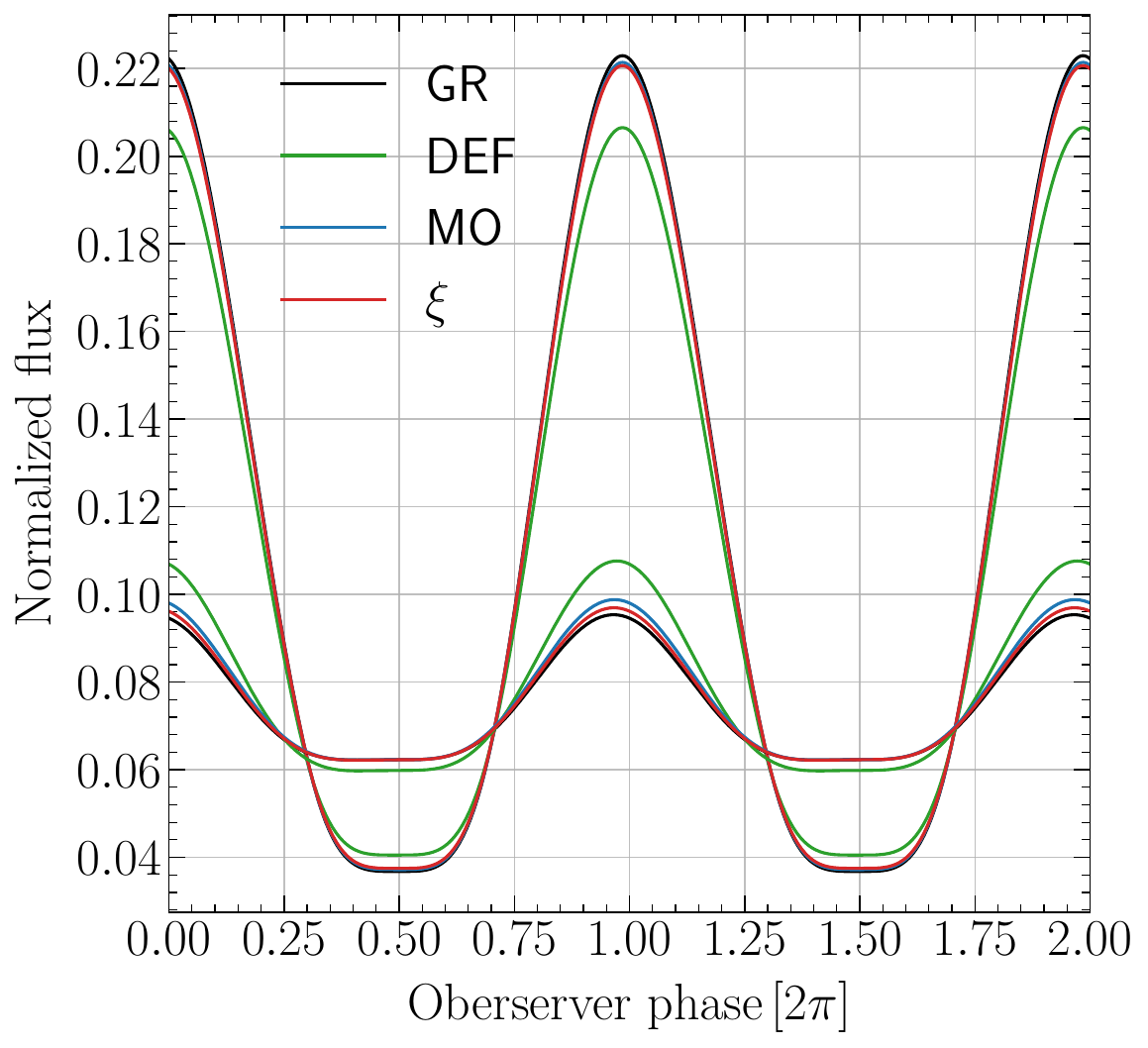}
  \caption{X-ray pulse profiles for NSs with anisotropic emission intensity
  $I^{\prime}$. Here the curves are results from point-like hot spots. The
  masses of the NSs are chosen as $1\,M_{\odot}$ for the upper set of curves and
  $2\,M_{\odot}$ for the lower set of curves. The integral radiating intensities
  of hot spots on the surface of NSs are chosen to be the same. Other parameters
  are the same as in Fig.~\ref{fig:T-F}. We have used EOS AP4 in the
  calculation.  \label{fig:T-F_ld} }
\end{figure}

In the above discussion, the specific intensity of the radiation of the two
hot spots is assumed to be isotropic. This assumption holds if the two hot
spots are blackbodies and sit in vacuum. However, for realistic NSs, the
presence of magnetic fields and Compton scattering can result in the
``limb-darkening'' effect so that the radiation intensity decreases with the
angle $\alpha'$ ~\cite{Zavlin:1996wd, Bogdanov:2006zd}. This effect can change
the features of the profiles significantly. To our best knowledge,  this effect
is considered numerically in ST theories for the first time in this work. 

For numerical study of the limb-darkening effect, we take a simple form of the
intensity
\begin{equation}
  I'(\nu',\alpha')=\cos \alpha'\,.
\end{equation}
In Fig.~\ref{fig:T-F_ld} we show the X-ray profiles for the two NSs with the
same parameters as in Fig.~\ref{fig:T-F} but with the anisotropic specific
intensity. One can notice that the limb-darkening effect partly reduces the
effect of beaming and gives a much smoother curve. For NSs with spontaneous
scalarization, the main deviations from GR still occur at the peak of the
profiles.

To give a quantitative description of the differences in the X-ray profiles
between GR and the ST theories, we present some numerical results in Table
\ref{tab:x-ray}. In the table, we show the peak values of the normalized fluxes
and the related observer phases for different choices of parameters. One can see
that the deviations are mainly in the peak fluxes of the profiles, which can be
about 10\% for the DEF theory and about 2\% for the $\xi$ and the MO theories.
The deviations in phase are generally small, and can actually only be seen in
the case of $M=2\,M_{\odot}$. The deviations for the DEF theory are much larger
than those of the $\xi$ and the MO theories, which is consistent with the size
of scalarization shown in Fig.~\ref{fig:phic-M}.  One may notice that for the last
column of the DEF theory, the peak fluxes and the deviations are smaller than
those of the MO and the $\xi$ theory. This is because the DEF theory has strong
scalarization so the radius of the NS is larger compared to the other two
theories, and in this case, the radius is too large so that the flux from one of
the hot spots is blocked by the star itself when the total flux reaches the
peak. For the MO or the $\xi$ theories, the scalarization is weaker than the DEF
theory and the radius of the NS of the same mass is smaller.  The fluxes from
both of the hot spots can be received by the observer when the total flux
reaches the peak.

\section{Tidal Love numbers in the ST theories}
\label{sec:tidal}

GW observation of binary NS mergers provides unique possibilities to study the
EOS of the NS and to test GR ~\cite{Hinderer:2007mb,Damour:2009vw}. At the early
stage of the inspiral, the evolution of the system is dominated by the point-mass
dynamics. But as the system evolves to the end of the inspiral, the finite-size
effect starts to show its influence on the GW signal. Phenomena like tidal
effects which depend on the EOS of the NS can lead to a small change of the 
waveform's phasing behaviour. Thanks to the matched filtering technique used in
the signal analysis, high sensitivities to the phase evolution in GWs can be
achieved, therefore leading to measurements of tidal effects. The well-known GW
discovery of  GW170817 has been used to constrain the tidal deformability
parameter for the first time~\cite{LIGOScientific:2017vwq,
LIGOScientific:2018hze}.

NSs' tidal deformability is defined as the ratio of the induced tidal
deformation to the strength of the tidal perturbation~\cite{Hinderer:2007mb}. In
GR, the metric coefficient $g_{tt}$ of a tidally-deformed NS can be expanded as
\begin{eqnarray}
  -\frac{1+g_{tt}}{2}&=&-\frac{M}{\rho} -\frac{3 Q_{ij}}{2 \rho^3}\big(n^i n^j-\frac{1}{3}\delta^{ij}\big)
  +O\Big(\frac{1}{\rho^4}\Big)\nonumber\\
  &&+\frac{1}{2}{\cal E}_{ij} x^i x^j +O\big(\rho^3\big)\,,\label{eqn:gtt}
\end{eqnarray} 
in the star's local
asymptotic rest frame~\cite{Hinderer:2007mb}, where $M$ is the star's mass, $Q_{ij}$ is the
induced quadrupole moment, ${\cal E}_{ij}$ is the external tidal field, and 
$n^i \equiv x^i/\rho$. To the linear order in ${\cal E}_{ij}$, the induced
quadrupole will have the form $Q_{ij}=-\lambda {\cal E}_{ij}$.
Here, the parameter $\lambda$ represents the tidal deformability of the star, which is related to the $l=2$ 
tidal Love number $k_{2}$ via $k_2={3}G\lambda R^5/2$.

Considering ST theories, we need to expand the metric with the possible
perturbations in the scalar field. For the massive scalar field, the asymptotic
behavior of the scalar field and its perturbation have the form
\begin{equation}
  \varphi ~ \mbox{and} ~ \delta\varphi  \quad \to \quad \frac{1}{\rho}e^{-\rho/\lambda_{\varphi}}\,.
\end{equation}
Such an exponential drop will not affect the expansion (\ref{eqn:gtt}) in any
order of $1/\rho$, which is different from the massless
case~\cite{Pani:2014jra}. \Reply{}{Even though the scalar field perturbation 
is Yukawa-suppressed, the scalarized NS itself will respond 
differently to a tidal field, which leads to a different 
tidal deformability.}

Restricting to the $l=2$, static, even-parity perturbations in the Regge-Wheeler gauge~\cite{Regge:1957td}, 
the perturbations in the metric and the scalar field can be written as
\begin{eqnarray}
  \delta g_{\mu\nu}^{(2m)}&=&Y_{2m}(\theta,\phi)\nonumber
  \left[ 
    \begin{matrix}
      -e^{\nu}H_0 & H_1 & 0 & 0 \\
      H_1 & H_2/\left( 1-\frac{2m}{\rho}\right) & 0 & 0 \\
      0 & 0 & \rho^2K & 0 \\
      0 & 0 & 0 & \rho^2 \sin^2\theta K
    \end{matrix}
  \right]\,,\nonumber\label{eqn:deltag}\\
  \\
  \delta \varphi^{(2m)}&=& Y_{2m}(\theta,\phi)\, \delta \varphi \, ,\label{eqn:deltaphi}
\end{eqnarray}
where $H_0\,,H_1\,,H_2\,,K\, $ and $\delta \varphi$ are functions only depending
on $\rho$, and $Y_{lm}$ are the spherical harmonics.

Following the procedure in the original work by \citet{Hinderer:2007mb}, we
combine Eqs.~(\ref{eqn:deltag}--\ref{eqn:deltaphi}) with the linearized Einstein
equations, and get 
\begin{eqnarray}
  H_1&=&0\,,\\
  H_0&=&-H_2=H(\rho)\,,\\
  K&=&\int \left[-H'(\rho)-H(\rho)\nu'(\rho) -4\delta \varphi(\rho)\varphi(\rho)\right] \rmd \rho\,.
\end{eqnarray}
Together with the linearized perturbation of Eq.~(\ref{eqn:phi}), one finally
gets the differential equations for $H$ and $\delta\varphi$,
\begin{eqnarray}
  \label{eqn:H}
  H''+c_1H'+c_0H&=&c_s\delta \varphi\,,\\
  \label{eqn:delphi}
  \delta \varphi'' +d_1 \delta \varphi' +d_0 \delta \varphi &=&d_sH\,,
\end{eqnarray}
where 
\begin{widetext}
\begin{eqnarray}
  c_1&=&d_1=\frac{4-2x-\rho^2V+8\pi \rho^2 A^4 (\tilde{p}-\tilde{\epsilon})}{2\rho(1-x)}\,,\\
  c_s&=&4d_s=\frac{1}{1-x}\Bigg\{8\pi  A^4 \alpha \bigg[\tilde{\epsilon}\Big(1-\frac{\rmd \tilde{\epsilon}}{\rmd \tilde{p}}\Big) + \tilde{p}\Big(9-\frac{\rmd \tilde{\epsilon}}{\rmd \tilde{p}}\Big)\bigg]+32\pi \rho A^4 \tilde{p}\varphi' - 
  \frac{\rmd V}{\rmd \varphi} - 2\rho \varphi'\bigg[V-2\varphi'^2+  2x\Big(\varphi'^2- \frac{1}{\rho^2}\Big)\bigg]\Bigg\}\,,\\
  c_0&=&\frac{4\pi A^4 (\tilde{p}+\tilde{\epsilon})}{1-x} \frac{\rmd
  \tilde{\epsilon}}{\rmd \tilde{p}} - \frac{1}{
  (1-x)^2}\Bigg\{ \frac{x^2}{\rho^2} (\rho^2\varphi'^2-1)^2 - x
  \bigg[ \frac{6}{\rho^2} -2 \varphi'^2+2\rho^2\varphi'^4+V(2-\rho^2\varphi'^2)
  +4\pi A^4\Big(\tilde{p}\big[4\rho^2\varphi'^2 -13\big] -5\tilde{\epsilon}
  \Big)\bigg] \nonumber\\
  && +
  64\pi^2\rho^2A^8\tilde{p}^2+ \frac{\rho^2V^2}{4} + \frac{6}{\rho^2}+\rho^2\varphi'^4 +  V(1-\rho^2\varphi'^2)
  - 4\pi A^4 \Big[5\tilde{\epsilon}+\tilde{p}\big(9+2\rho^2V -
  4\rho^2\varphi'^2\big)\Big] \Bigg\}\,,\\
  d_0&=&\frac{4}{1-x}\Bigg\{\pi A^4\alpha^2(\tilde{p}+\tilde{\epsilon})
  \frac{\rmd \tilde{\epsilon}}{\rmd \tilde{p}} + \bigg[
  6\pi A^4\alpha^2(\tilde{p}-\tilde{\epsilon})- (1-x)\varphi'^2 -\frac{3}{2\rho^2} \bigg] +
  \pi  A^3 (3\tilde{p}-\tilde{\epsilon}) \frac{\rmd^2 A}{\rmd \varphi^2}  -\frac{1}{16}  \frac{\rmd^2 V}{\rmd \varphi^2} \Bigg\}\,,  
\end{eqnarray}
\end{widetext}
where $x \equiv 2m/\rho$.

\begin{figure*}
  \centering 
  \includegraphics[width=12cm]{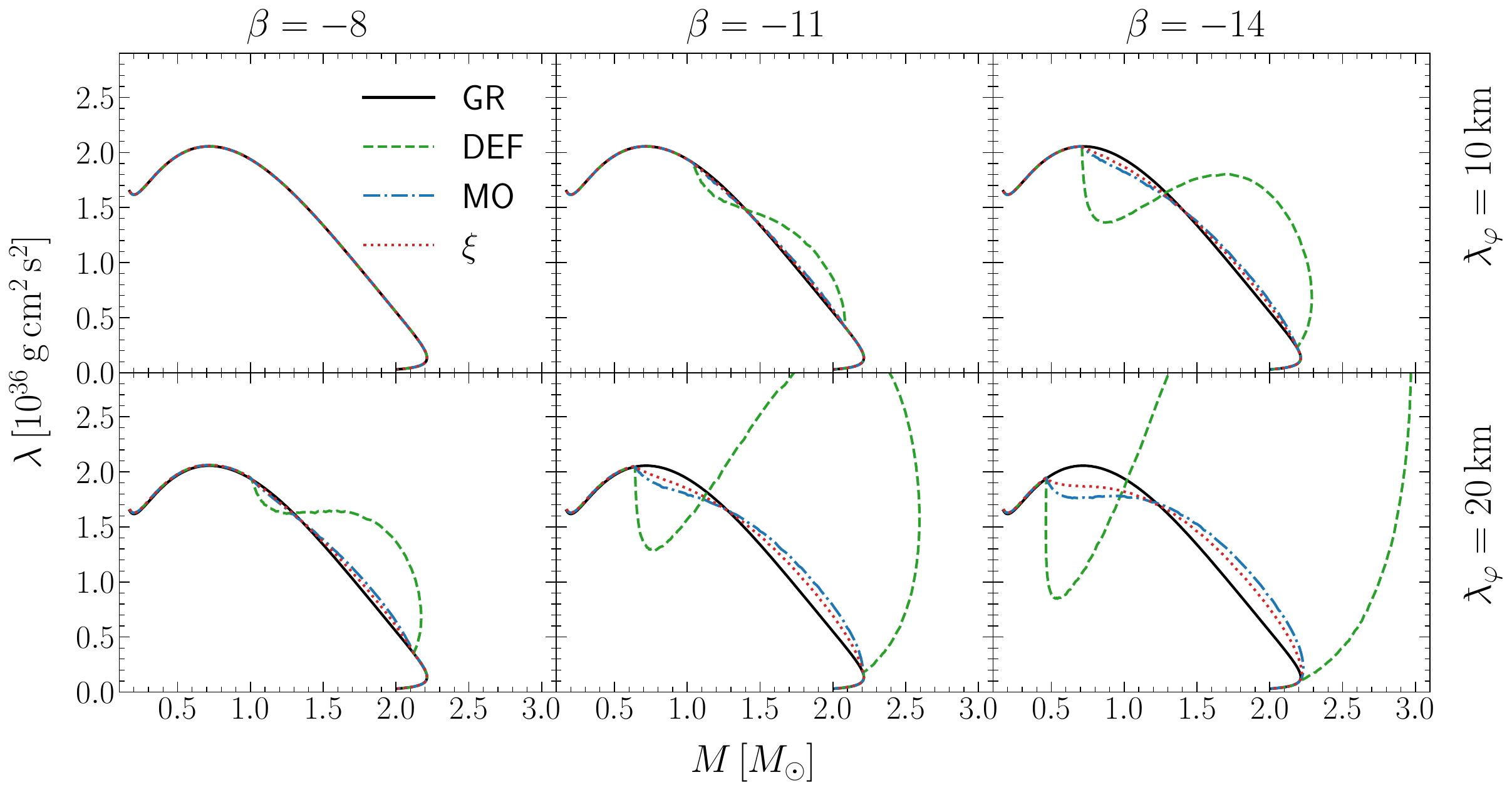}
  \caption{The tidal deformability $\lambda$ versus the ADM mass $M$ in different ST theories
  with EOS AP4.}
  \label{fig:M-lambda}
\end{figure*}

\begin{table*}
  \renewcommand\arraystretch{1.3}
  \begin{center}
    \caption{Tidal deformabilities and their percentage deviations from GR in the ST
    theories with $M=1\,{M_{\odot}}$ and $M=2\,{M_{\odot}}$. We have used EOS
    AP4 in the calculation.}
    \begin{tabular*}{\hsize}{@{}@{\extracolsep{\fill}}ccm{0.05\textwidth}<{\centering} m{0.05\textwidth}<{\centering}m{0.05\textwidth}<{\centering}m{0.05\textwidth}<{\centering} m{0.05\textwidth}<{\centering}m{0.05\textwidth}<{\centering}m{0.05\textwidth}<{\centering} m{0.05\textwidth}<{\centering}m{0.05\textwidth}<{\centering}m{0.05\textwidth}<{\centering} m{0.05\textwidth}<{\centering}m{0.05\textwidth}<{\centering}m{0.05\textwidth}<{\centering} @{}}
      \hline\hline
      \multicolumn{2}{c}{\multirow{2}{*}{Theory}}&\multirow{2}{*}{GR}&\multicolumn{3}{c}{$\beta=-11,\,\,{\lambdabar_{\varphi}}=10\,\rm km$}&\multicolumn{3}{c}{$\beta=-14,\,\,{\lambdabar_{\varphi}}=10\,\rm km$}&\multicolumn{3}{c}{$\beta=-11,\,\,{\lambdabar_{\varphi}}=20\,\rm km$}&\multicolumn{3}{c}{$\beta=-14,\,\,{\lambdabar_{\varphi}}=20\,\rm km$}\\
      &&&DEF& MO & $\xi$&DEF&MO&$\xi$&DEF&MO&$\xi$&DEF&MO&$\xi$\\
      \hline
      \multirow{2}{*}{$M=1\,M_{\odot}$}&$\lambda \, [\rm 10^{36}\,g\,cm^2\,s^2]$&1.94&1.94&1.94&1.94&1.41&1.82&1.87&1.57&1.79&1.85&1.85&1.78&1.82\\
      &deviation (\%)&0&0.00&0.00&0.00&27.0&5.97&3.58&19.1&7.59&4.64&4.37&8.26&6.09\\
      \hline
      \multirow{2}{*}{$M=2\,M_{\odot}$}&$\lambda \, [\rm 10^{36}\,g\,cm^2\,s^2]$&0.551&0.857&0.578&0.567&1.62&0.644&0.609&3.17&0.786&0.693&5.29&0.865&0.757\\
      &deviation (\%)&0&55.5&4.90&2.92&194&16.8&10.5&475&42.6&25.7&860&56.9&37.3\\
      \hline
    \end{tabular*}\label{tab:lambda}
  \end{center}
\end{table*}

Note that Eqs.~(\ref{eqn:H}--\ref{eqn:delphi}) are linear equations of $H$ 
and $\delta\varphi$. To solve them, one can integrate the system twice with the 
initial values~\cite{Pani:2014jra} 
\begin{equation}
  H|_{R_0}=R_0^2\,,\ H'|_{R_0}=2R_0\,,\ \delta\varphi|_{R_0}=0\,,\ \delta\varphi'|_{R_0}=0\,,
\end{equation}
and
\begin{equation}
  H|_{R_0}=0\,,\ H'|_{R_0}=0\,,\ \delta\varphi|_{R_0}=R_0^2\,,\
  \delta\varphi'|_{R_0}=2R_0 \,,
\end{equation}
respectively, where $R_0$ is a small radius. Then one needs to make a linear
combination of these two results to construct a solution whose asymptotic value
of $\delta \varphi$ vanishes.

For a massive scalar field, the asymptotic form of the metric is the same as that in GR.
So one can integrate the system to a sufficiently large $\rho=\rho_i$ and ignore
the scalar field for $\rho>\rho_i$.  Define 
\begin{equation}
  C= \left. \frac{m}{\rho}\right|_{\rho\to\rho_i}\,, \quad  y= \left.\frac{\rho H'}{H}\right|_{\rho\to\rho_i}\, ,
\end{equation}
and then the tidal deformability can be calculated by~\cite{Hinderer:2007mb}
\begin{eqnarray}
  \lambda =&& \frac{2}{3} \rho_i^5 \frac{8C^5}{5} (1-2C)^2 \Big[2+2C(y-1)-y \Big]\nonumber\\
  &\times& \bigg\{ 4C^3 \Big[ 13-11y +C(3y-2)+2C^2(1+y) \Big] \nonumber\\
  &&+3(1-2C)^2 \Big[ 2-y+2C(y-1) \Big] \ln (1-2C) \nonumber\\
  && + 2C \Big[ 6-3y+3C(5y-8) \Big] \bigg\}^{-1}\,.
\end{eqnarray}

\Reply{}{The case is different for a massless scalar field, as can be seen in 
Ref.~\cite{Pani:2014jra}. The nonzero scalar charge developed by the scalar field 
will affect $\lambda$ and make it different in the Einstein and physical frame; see 
Eq.~(B8) in Ref.~\cite{Pani:2014jra}. }

In Fig.~\ref{fig:M-lambda}, we show the relation between the tidal deformability
$\lambda$ and the NS's ADM mass $M$ for the ST theories and GR. The parameters for the
ST theories are the same as those in Fig.~\ref{fig:phic-M}. Since the tidal
deformability is proportional to $R^5$, the deviations of the tidal
deformability follow the trend of, but are qualitatively larger than, the mass-radius
relation shown in Fig.~\ref{fig:M-R-P}. 

From Fig.~\ref{fig:M-lambda} we can see that the largest deviation happens
around $M=1\,{M_{\odot}}$ and $M=2\,{M_{\odot}}$, so we list the deviations of
the tidal deformability in the ST theories with $M=1\,{M_{\odot}}$ and
$M=2\,{M_{\odot}}$ in Table~\ref{tab:lambda}. The deviations increase quickly
with $|\beta|$ and $\lambdabar_{\varphi}$. The scalarized NS with
$M=2\,{M_{\odot}}$ in the DEF theory can have a tidal deformability almost 10
times greater than that in GR, thus providing an important window for tests.

\section{Universal relation in massive scalar-tensor theories}
\label{sec:I-lambda}

\begin{figure*}
  \centering 
  \includegraphics[width=14cm]{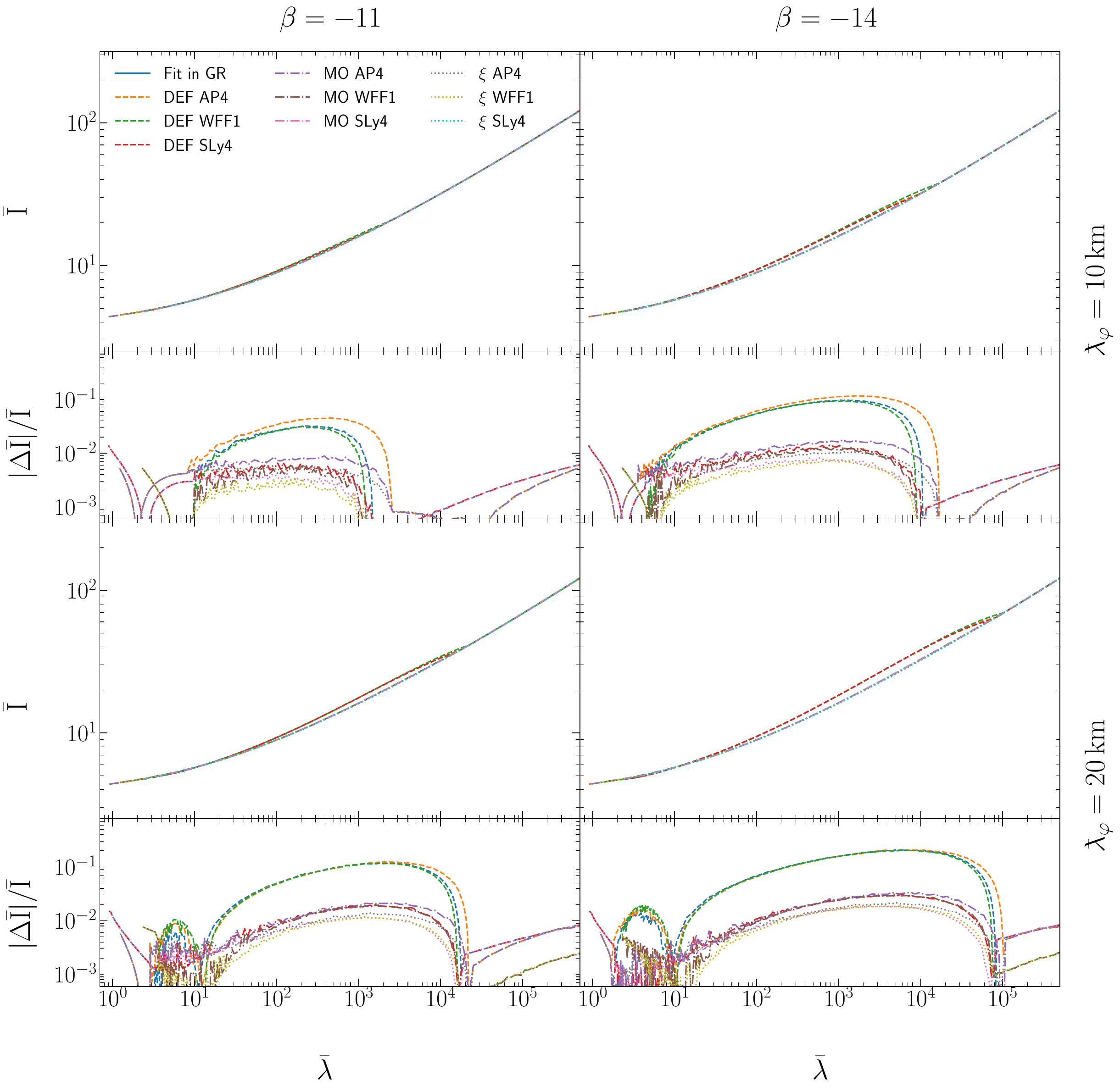}
  \caption{$\bar{I}$-$\bar{\lambda}$ universal relation for three ST theories 
  with different $\beta$ and ${\lambdabar_{\varphi}}$. We consider their EOSs,
  i.e., AP4, WFF1, and SLy4. The fitted curve for GR is taken from
  Ref.~\cite{Yagi:2016bkt}.}
  \label{fig:I-lambda}
\end{figure*}

The dimensionless moment of inertia $\bar{I}={I}/{M^3}$ and the dimensionless
tidal Love number  $\bar{\lambda}={\lambda}/{M^5}$ satisfy a remarkable
universal relation in GR \citep{Yagi:2013bca,Yagi:2013awa}. This relation is not
affected by the NS EOS up to an accuracy of a few percents. It will allow people
to test GR without knowing the details of the NS EOS, if the universal relations
in  ST theories are sufficiently different from those in GR.
\citet{Pani:2014jra} have calculated the slowly rotating and tidally deformed
NSs in the massless DEF theory. They find that the $\bar I$-$\bar \lambda$-$\bar
Q$ relations agree with GR in a high accuracy for the allowed parameter space
$\beta>-4.5$, which is set by the timing data of the pulsar--white dwarf
systems~\citep{Shao:2017gwu,  Anderson:2019eay, Zhao:2019suc, Guo:2021leu}. But
the deviation can reach to nearly 10\% for some parameters that are already
ruled out.  \citet{Doneva:2014faa} calculated the $\bar{I}$-$\bar{Q}$ in the
massless DEF theory for rapidly rotating NSs with $\beta=-4.5$.  The universal
relation between $\bar I$ and $\bar Q$ still holds and the deviations from GR
are small and hard to test with future astrophysical observations.

Due to the Yukawa suppression originated from the mass term, massive ST theories
can avoid most of the constraints of previous observations. Thus one might
expect considerable deviations from GR in the universal relations in certain
yet-allowed parameter space. \citet{Doneva:2016xmf} constructed the model of
rapidly rotating NSs in the massive DEF theory. They find that the
$\bar{I}$-$\bar{Q}$ universal relation is nearly EOS independent and the
deviations from GR can be large for some allowed parameters.

In Fig.~\ref{fig:I-lambda} we plot the $\bar{I}$-$\bar{\lambda}$ universal
relation for the massive ST theories with different choices of parameters. 
Panels in the first and third rows are the $\bar{I}$-$\bar{\lambda}$ relation
for EOSs AP4~\cite{Akmal:1998cf}, WFF1~\cite{Wiringa:1988tp}, and
SLy4~\cite{Douchin:2001sv}. Panels in the second and fourth rows are the
relative deviations from the fitted curve, which is the GR result given in
Ref.~\cite{Yagi:2016bkt}. From the figure we can see that the
$\bar{I}$-$\bar{\lambda}$ universal relation still holds in the massive ST
theories to a rather good precision. For the MO theory and the $\xi$-theory, the
largest deviation from GR is within only a few percents. Though the deviations
increase with $|\beta|$ and $\lambdabar_\varphi$, there exist upper limits for
the coupling function in these two theories as shown in
Fig.~\ref{fig:alpha-phi}, and thus the deviations in the
$\bar{I}$-$\bar{\lambda}$ relation also stay finite and not too large.
Differently for the DEF theory, the coupling function shown in
Fig.~\ref{fig:alpha-phi} is unbound as $\varphi$ increases, so the deviation in
the $\bar{I}$-$\bar{\lambda}$ relation can be as large as 10\% in
Fig.~\ref{fig:I-lambda}. Comparably significant deviations in the
$\bar{I}$-$\bar{\lambda}$ relation as the scalarization becomes strong were also
observed in the massless DEF theory in Ref.~\cite{Pani:2014jra}.

\begin{figure}
  \centering 
  \includegraphics[width=0.9\linewidth]{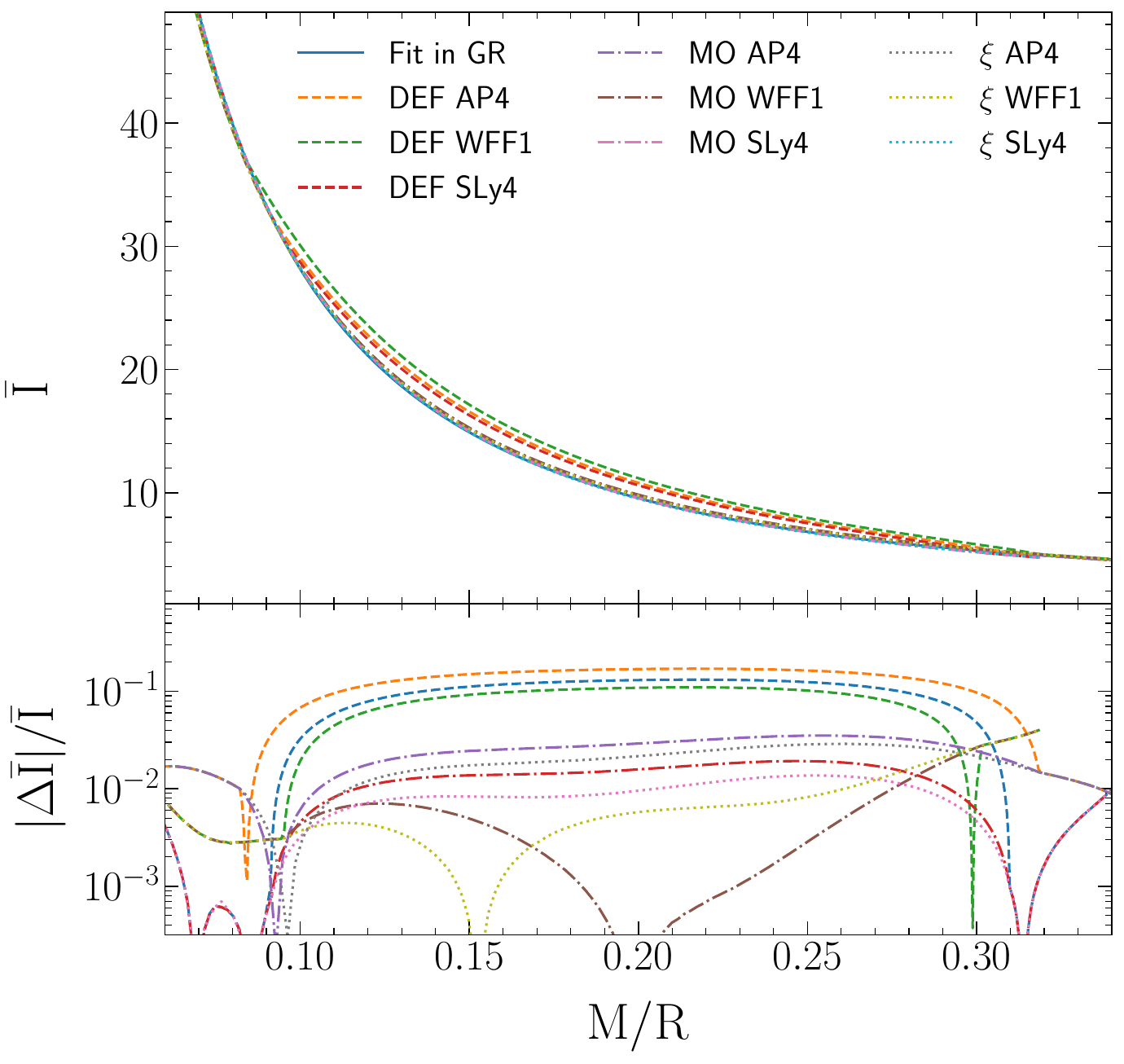}
  \caption{The universal relation between $\bar{I}$ and $M/R$ for three 
  ST theories with $\beta=-14$ and ${\lambdabar_{\varphi}=10\, \rm km}$. The 
  fitted curve is derived by a fifth polynomial fitting for GR with 
  EOSs AP4, WFF1, and SLy4. \label{fig:C-I} }
\end{figure}

\Reply{}{We mention that there is also a universal relation between the 
dimensionless moment of inertia $\bar{I}$ and the star's compactness $C = M/R$ 
in both GR~\cite{Breu:2016ufb} and scalar tensor theories~\cite{Minamitsuji:2016hkk}, 
as show in Fig.~\ref{fig:C-I}. We show the relation in three ST theories with 
EOSs AP4, WFF1, and SLy4 for $\beta=-14$ and ${\lambdabar_{\varphi}=10\, \rm km}$. 
The fitted curve in GR is derived with a fifth polynomial fitting. The relation 
between the moment of inertia and the star's compactness seems to have larger 
dependence on the EOS than the $\bar{I}$-$\bar{\lambda}$ universal relation.}

\section{Summary}
\label{sec:summary}

In this work, we study the NS structures in the DEF, the MO and the $\xi$
theories with a massive scalar field. The coupling functions to matters differ
in these theories. Our investigation serves complementary studies to the
massless case. Differential equations for a slowly rotating NS are derived and
solved numerically. We analyze and compare the mass-radius relation and the
moment of inertia of scalarized NSs in those theories in detail, pertaining to
experiments like NICER~\cite{Watts:2016uzu,Miller:2021qha,Raaijmakers:2021uju}
and SKA~\cite{Hu:2020ubl}. The structures of the scalarized NSs in those
theories have qualitatively similar behaviours, except that the DEF theory, with
a linear coupling function, has much lager scalarization compared to the MO and
the $\xi$ theories whose coupling functions are bounded (see
Fig.~\ref{fig:alpha-phi}). Compared to the massless case, a massive scalar field
results in Yukawa suppression which in general introduces smaller deviations
from GR and thus can avoid current experimental constraints more easily.

As an application of the numerical results of the NS solutions, we calculate the
X-ray profiles from a pair of hot spots on the surfaces of scalarized NSs. We
consider a slowly rotating NS but neglect the frame-dragging effect in the
metric which is minute here anyway. By taking a model of extended hot spots,
which is done for the first time in ST theories, we show that for finite
circular hot spots with $1\, {\rm km}$ radius, the simple model of point-like
hot spots is still a good approximation. Deviations from GR mainly happen at the
peaks of the profiles.  For the DEF theory, the flux deviations can exceed 10\%
while for the MO and the $\xi$ theories, this number is about 3\%. The phase
deviations in the three  ST theories are all within 1\% which means that it is
hard to test with observations from pulsar timing. We also study the
limb-darkening effect and show that it can smooth the pulse profiles but not
much affect the level of deviations from GR. These studies will provide useful
inputs to X-ray timing missions like NICER in testing alternative gravity
theories.

In addition, for the first time we calculate the tidal deformability of NSs in
the massive ST theories. The results are qualitatively consistent with the
massless case, but due to the mass term of the scalar field, massive ST theories
have much larger yet-allowed parameter space unconstrained where large
deviations from GR are possible. We investigate the $\bar{I}$-$\bar{\lambda}$
universal relation in the massive ST theories. For the MO and the $\xi$ 
theories, the deviations from GR are within a few percents due to their finite
coupling functions; but for the DEF theory, the deviation can exceed 10\% and be
even higher for sufficiently large $|\beta|$ and $|\lambda_\varphi|$. It will be
interesting to test the massive ST theories with the universal relation via the
observations of binary NS mergers and their associated GWs.

\acknowledgments

We are grateful to Norbert Wex for carefully reading the manuscript and providing helpful comments, \Reply{}{and the anonymous referee for constructive suggestions}.
This work was supported by the National SKA Program of China (2020SKA0120300),
the National Natural Science Foundation of China (11975027, 11991053, 11721303),
the Young Elite Scientists Sponsorship Program by the China Association for
Science and Technology (2018QNRC001), the Max Planck Partner Group Program
funded by the Max Planck Society, and the High-performance Computing Platform of
Peking University.
ZH is supported by the Principal's Fund for the Undergraduate Student Research
Study at Peking University, and RX is supported by the Boya Postdoctoral
Fellowship at Peking University.

\bibliography{refs}

\end{document}